\documentclass[11pt,twoside,letterpaper,preprintnumbers]{revtex4}
\usepackage[dvips]{graphicx}
\begin{document}
\preprint{UH511-1009-02}
\title{Mass predictions based on a supersymmetric SU(5) fixed point}
\author{Javier Ferrandis}
\email{javier@phys.hawaii.edu}
\homepage{www.phys.hawaii.edu/~javier}
\affiliation{Department of Physics \& Astronomy\\
 University of Hawaii at Manoa\\
 2505 Correa Road\\
 Honolulu, Hawaii, 96822}
\begin{abstract}
I examine 
the possibility that 
the third generation fermion masses 
are determined
by an exact fixed point of the minimal supersymmetric
SU(5) model.
When one-loop supersymmetric thresholds are included,
this unified fixed point successfully predicts the top quark mass, 
$175 \pm 2$~GeV, as well as the weak mixing angle.
The bottom quark mass prediction is sensitive
to the supersymmetric thresholds; it approaches the
measured value for $\mu <0$ and very large
unified gaugino mass. 
The experimental measurement
of the tau lepton mass 
determines $\tan\beta$,
and the strong gauge coupling and fine structure 
constant fix the unification scale and the unified
gauge coupling.
\end{abstract}
\maketitle
\newpage
\section{Introduction
\label{sec1}}
%
The measurement of the top quark mass at Fermilab in 1995
\cite{Groom:in}
completed a search that started nearly a hundred
years ago with the measurement of the electron mass.
Although neutrino masses still remain to be accurately measured, 
we already have 
the most important pieces of the puzzle of fermion masses.
A satisfactory theory to explain the standard model (SM) fermion masses, however, 
is still lacking. 
The disparity in the mass scales of the third generation fermions
and the first and second generation fermions, as well as their mixing angles, 
suggest that different
mechanisms may be responsible for the masses of third generation,
and first and second generation fermions. Indeed, an explanation
of third generation fermion masses may be necessary to understand
first and second generation fermion masses. 

It is widely hoped that fermion masses and mixing angles 
will be explained, or at least reduced, 
by some grand unified theory (GUT)\cite{Georgi:sy} . In this case, 
to obtain predictions for the values of masses, gauge 
couplings and mixing angles,
the predictions for the boundary conditions of the unified theory at the GUT scale
would be extended, using the renormalization group equations,
from the unification scale down to the electroweak scale. 
The current grand unified theories, however,  
reduce the number of parameters but they do not make 
definite predictions for the Yukawa couplings at the unification scale. 
These couplings must be fit to the experimental measurements.
This approach has been extensively used in the literature
to study third generation fermion masses in the context of 
supersymmetric grand unified models (SUSY GUTs)
\cite{susyguts}.
It has been found that the unification scale 
Yukawa couplings can be adjusted to fit the measured 
top, bottom and tau fermion masses~\cite{SO10,texturas}.

There is an alternative idea,
not incompatible with the previous approach,
which can make unified theories more predictive.
If the renormalization
group equations describing the evolution of the
various couplings in the theory possess an infrared fixed
point, then some of these couplings could be swept towards
this fixed point \cite{Pendleton:as,masso}.
M.~Lanzagorta and G.G.~Ross \cite{Lanzagorta:1995gp} have pointed out
that the fixed point
structure of the unified theory beyond the standard model
may play a very important role in the determination
of the Yukawa couplings at the gauge unification scale.
It was argued 
that the evolution towards the fixed point may occur more
rapidly in some unified theories than in the low-energy
theory due to their larger field content and 
the possible lack of asymptotic freedom.
Therefore, even though
the ``distance'' between the grand unification scale
and the Planck scale (or the compactification scale)
is much less than the distance between the
electroweak scale and the grand unification scale,
the unified fixed point could play a significant role
in determining the couplings at the unification scale,
regardless of their values in the underlying ``Theory of
Everything." 
This possibility would make the 
unified theory much more predictive because
the Yukawa couplings at the unification scale may be determined 
in terms of the unified gauge
coupling according to the unified gauge group
and the multiplet content of the model. 

The purpose of this paper is to study the possibility that 
the third generation fermion masses are determined
by the fixed point structure of the minimal supersymmetric
SU(5) model. 
We will not analyze here whether or not nature can reach a fixed
point at the unification scale, 
because ultimately that analysis depends
on assumptions about the still unknown 
physics at yet higher energy scales, perhaps at the Planck scale.
Instead, we will adopt a phenomenological approach: 
we will assume that at the unification scale the SU(5)
model is at a fixed point, and we will study the
low-energy implications of 
our assumption. 

We will not address here the problems that afflict 
the minimal supersymmetric SU(5) model: namely, 
the doublet-triplet splitting problem,
the strong experimental constraints on proton decay,
the generation of neutrino masses, and the lack of predictivity
in the flavor sector.
Solutions have been proposed for each of these problems,  and entail 
complicating the minimal SU(5) model \cite{Masina:2001pp}.
Since our aim is to explore a new idea,
in this paper we will use
the smallest simple group in which one can embed the
standard model gauge group:
the minimal third-generation supersymmetric SU(5) model.
We think that the idea can be extended to non minimal 
grand unified models, where the problems that afflict 
minimal SU(5) may find a resolution.
 
This paper is organized as follows. 
We begin in Sec.~\ref{sec2} with the calculation of the fixed points for the 
Yukawa sector of the model. We find only one 
phenomenologically interesting fixed point and analyze 
its mass predictions, first by
neglecting low-energy supersymmetric thresholds. In Sec.~\ref{sec3}
we calculate the associated fixed point in the soft SUSY breaking sector
of the model. In Sec.~\ref{sec4} we study the fixed point
implications for the breaking of the SU(5) symmetry.
In Sec.~\ref{sec5} we study the characteristic 
low-energy supersymmetric spectra predicted 
by the fixed point. 
In Sec.~\ref{sec6} we study the fermion mass predictions
including the low-energy supersymmetric thresholds. Here, we also analyze  
the dependence of the predictions on the unified gaugino mass and on 
the sign of the $\mu$-term.
In Sec.~\ref{sec7} we briefly analyze the fixed point
predictions for fermion masses 
in the context of non minimal SU(5) models with additional
particle content. We conclude in Sec.~\ref{sec8} with a brief summary
of our results.
%
\section{Fixed points for the minimal supersymmetric SU(5)
\label{sec2}}
%
We begin our calculation of 
one-loop fixed points 
with a brief review of the model
to set up our conventions and notation.
The superpotential of the 
minimal one-generation supersymmetric SU(5) model, 
omitting SU(5) indices, is given by 
\cite{Witten:nf,Sakai:1981gr,Dimopoulos:1981zb}
\begin{eqnarray}
{\cal W}_{\rm SU(5)} &=& \mu_{\Sigma}~ tr \Sigma^2 + \frac{1}{6}\lambda_{\Sigma}~ 
tr \Sigma^3 + \mu_{H}~ {\cal H}_{5} {\cal H}_{\overline{5}}  
\nonumber \\
&+&\lambda_{H} {\cal H}_{\overline{5}}\Sigma {\cal H}_{5}
+ \frac{1}{4} \lambda_t \psi_{10} \psi_{10}  {\cal H}_{5} + 
\sqrt{2} \lambda_b \psi_{10} \psi_{\overline{5}} {\cal H}_{\overline{5}},
\end{eqnarray}
where $\psi_{10}$ and $\psi_{\overline{5}}$ are matter chiral superfields
belonging to representations {\bf 10} and $\overline{{\bf 5}}$
of SU(5), respectively.
As in the supersymmetric generalization of the standard model,
to generate fermion masses
we need two sets of Higgs superfields,  ${\cal H}_{5}$
and ${\cal H}_{\overline{5}}$, belonging to representations {\bf 5} and 
$\overline{{\bf 5}}$ of SU(5).
The Higgs superfield
in the adjoint {\bf 24}-dimensional 
representation, $\Sigma$, is chosen to allow the existence of a realistic
breaking of the SU(5) symmetry down to the SM group, 
${\rm G}_{\rm SM} \equiv SU(3)_C \times SU(2)_L \times U(1)_Y$.
Typically, fixed points appear for ratios of couplings, 
such as  ratios of Yukawa couplings to gauge couplings,
and are known generically as Pendleton-Ross fixed points.
The renormalization group equations (RGEs)
for the SU(5) Yukawa couplings, valid above the GUT scale, are given by
\cite{pomarol,strumia,Bagger:1999ty}
%
\begin{equation}
\frac{d}{dt}
\left(
\begin{array}{c}
{\lambda}_t^2 \\
{\lambda}_b^2 \\
{\lambda}^2_{H} \\
{\lambda}^{2}_{\Sigma}
\end{array}
\right) =
\left(
\begin{array}{c}
{\lambda}_t^2 \\
{\lambda}_b^2 \\
{\lambda}^2_{H} \\
{\lambda}^{2}_{\Sigma} 
\end{array}
\right)^{T}
\left(
\left(
\begin{array}{cccc}
9& 4 & 24/5  & 0 \\
3& 10 & 24/5 & 0 \\
3 & 4 & 53/5 & 21/20 \\
0 & 0 & 3 & 63/20
\end{array}
\right)
\left(
\begin{array}{c}
{\lambda}_t^2 \\
{\lambda}_b^2 \\
{\lambda}^2_{H} \\
{\lambda}^{2}_{\Sigma} 
\end{array}
\right)
-
\left(
\begin{array}{c}
96/5 \\
84/5 \\
98/5 \\
30 
\end{array}
\right) g^2
\right),
\label{su5Yukrge}
\end{equation}
where $T$ denotes the matrix transpose, 
$t= 1/8 \pi^2  \log \Lambda$, and $\Lambda$ is 
the renormalization scale.
The RGE for the SU(5) unified gauge coupling is given by
%
\begin{equation}
\frac{dg^2}{dt}  =   b g^4,
\label{su5gaurge}
\end{equation}
with $b=-3$ in the minimal model.
The RGEs for the ratios of Yukawa couplings to the
unified gauge coupling can be easily calculated from Eqs.~(\ref{su5Yukrge})
and \ref{su5gaurge}:
\begin{equation}
\frac{1}{g^2}\frac{d}{dt}
\left(
\begin{array}{c}
\widetilde{\lambda}_t^2 \\
\widetilde{\lambda}_b^2 \\
\widetilde{\lambda}^2_{H} \\
\widetilde{\lambda}^{2}_{\Sigma} 
\end{array}
\right) =
 \left(
\begin{array}{c}
\widetilde{\lambda}_t^2 \\
\widetilde{\lambda}_b^2 \\
\widetilde{\lambda}^2_{H} \\
\widetilde{\lambda}^{2}_{\Sigma} 
\end{array}
\right)^{T}
\left(
\left(
\begin{array}{cccc}
9 & 4 & 24/5  & 0 \\
3 & 10 & 24/5 & 0 \\
3 & 4 & 53/5 & 21/20 \\
0 & 0 & 3 & 63/20
\end{array}
\right)
\left(
\begin{array}{c}
\widetilde{\lambda}_t^2 \\
\widetilde{\lambda}_b^2 \\
\widetilde{\lambda}^2_{H} \\
\widetilde{\lambda}^{2}_{\Sigma} 
\end{array}
\right)
-
\left(
\begin{array}{c}
96/5+b \\
84/5+b \\
98/5+b \\
30+b
\end{array}
\right)
\right),
\label{su5Yukgaurge}
\end{equation}
%
where we have defined
%
\begin{equation}
\widetilde{\lambda}_t^2 =
\frac{\lambda_t^2}{g^2},
\qquad
\widetilde{\lambda}_b^2 =
\frac{\lambda_b^2}{g^2},
\qquad
\widetilde{\lambda}^2_{H} =
\frac{\lambda^2_H}{g^2},
\qquad
\widetilde{\lambda}^{2} =
\frac{\lambda^{2}_{\Sigma}}{g^2}.
\label{yukrat}
\end{equation}
\begin{table} \centering    
\begin{tabular}{|c|c|c|c|c|} \hline \hline 
fixed point   & $\widetilde{\lambda}_t^2$ &  $\widetilde{\lambda}^2_b$ 
&  $\widetilde{\lambda}^2_H$ &  $\widetilde{\lambda}^2_{\Sigma}$  \\ \hline 
1 & $0~(R)$ & $0~(R)$ & $0~(R)$ & $0~(R)$ \\

2 & $0~(R)$ & $0~(R)$ & $0~(R)$ & $60/7~(A)$ \\
3 & $0~(R)$ & $0~(R)$ & $83/53~(A)$ & $0~(R)$ \\ 
4 & $0~(R)$ & $69/50~(A)$ & $0~(R)$ & $0~(R)$ \\ 
5 & $9/5~(A)$ & $0~(R)$ & $0~(R)$ & $0~(R)$ \\

6 & $0~(R)$ & $0~(R)$ & $19/24~(A)$ & $985/126~(A)$ \\
7 & $0~(R)$ & $69/50~(A)$ & $0~(R)$ & $60/7~(A)$ \\ 
8 & $9/5~(A)$ & $0~(R)$ & $0~(R)$ & $60/7~(A)$ \\ 
9 & $767/675~(A)$ & $0~(R)$ & $56/45~(A)$ & $0~(R)$ \\ 
10 & $0~(R)$ & $333/434~(A)$ & $277/217~(A)$ & $0~(R)$ \\ 

11 & $0~(R)$ & $5/4~(A)$ & $13/48~(A)$ & $2095/252~(A)$ \\
12 & $124/75~(A)$ & $0~(R)$ & $11/40~(A)$ & $349/42~(A)$ \\ 
13 & $89/65~(A)$ & $63/65~(A)$ & $0~(R)$ & $0~(R)$ \\
14 & $89/65~(A)$ & $63/65~(A)$ & $0~(A)$ & $60/7~(A)$ \\ 
15 & $132/95~(A)$ & $94/95~(A)$ & $-25/456~(R)$ & $20645/2394~(A)$ \\
16 & $2533/2605~(A)$ & $1491/2605~(A)$ & $560/521~(A)$ & $0~(R)$ \\ \hline \hline
\end{tabular} 
  \caption{\rm One-loop fixed points and their
stability properties for the minimal supersymmetric SU(5) model.}
  \label{su5fixedpoints}    
\end{table} 
We present in Table~\ref{su5fixedpoints} the list of
fixed points, along with their respective  solutions and 
stability character in the infrared \cite{Allanach:1997bh}, 
(A) if the direction is attractive, and (R) if it is repulsive. 
We observe that 12 out of the 16 solutions
have vanishing 
$\lambda_t$ or $\lambda_b$ and are phenomenologically
not viable, as they predict a zero mass for the
top quark, or for the bottom quark and tau lepton.
Moreover, it is known that the dimension-five operators induced by the
colored Higgs triplet give large contributions to the 
proton decay rate \cite{proton}.
To suppress such operators, the mass of the colored Higgs triplet
has to be heavy, implying $\lambda_H > g$. 
This then excludes solutions 13 and 14, which 
have $\lambda_H=0$.
Solution 15 is a non physical solution with $\lambda_H^2 <0$.
Only one interesting solution remains, solution 16, which 
predicts
%
\begin{equation}
\frac{\lambda_t^2}{g^2}  =   \frac{2533}{2605},
\qquad
\frac{\lambda_b^2}{g^2}  =   \frac{1491}{2605},
\qquad
\frac{\lambda^2_{H}}{g^2}  =   \frac{560}{521},
\qquad
\frac{\lambda^{2}_{\Sigma}}{g^2}  =  0.
\label{su5fpyuksol}
\end{equation}
We observe that this fixed point predicts 
$\lambda_{H} = 1.03675 ~g$, overcoming the previous na\" \i ve
constraint on proton decay.
Moreover, it predicts interesting values for the top
and bottom-tau Yukawa couplings at the unification scale,
$\lambda_t = 0.98608 ~g$ and $\lambda_b = 0.75654 ~g$
(later we will study, in detail, 
the fermion masses derived from
these predictions). 
On the other hand, the fixed point predicts $\lambda_{\Sigma}=0$,
while in principle we need $\lambda_{\Sigma} \neq 0$
if we want to break the SU(5) symmetry. 
Furthermore,
the directions $t$, $b$ and $H$ are attractive, 
while the direction $\Sigma$ is repulsive.
One can wonder to what extent it is consistent to study 
predictions coming from
this fixed point if the 
direction $\lambda_{\Sigma}=0$ is not attractive.
If we assume that the $\left| \lambda_{\Sigma} \right|$ 
coupling is much less than $g$,
the RGE can be approximately solved as
\begin{equation}
\frac{1}{g^2}\frac{d}{dt}
\widetilde{\lambda}^{2}_{\Sigma} \simeq
- 27~\widetilde{\lambda}^{2}_{\Sigma}
\longrightarrow
\widetilde{\lambda}^{2}_{\Sigma} (\Lambda_G)
\simeq \widetilde{\lambda}^{2}_{\Sigma}(\Lambda_P)
\times
\left(\frac{\Lambda_P}{\Lambda_G}\right)^{27/8\pi^2}
\simeq 
10~\widetilde{\lambda}^{2}_{\Sigma}(\Lambda_P).
\end{equation}
Then, $\lambda_{\Sigma}$ is only a slowly decreasing
function of the scale. For instance, assuming that 
$\widetilde{\lambda}_{\Sigma}^2 (\Lambda_{P}) < 10^{-3}$
at the Planck scale,
we obtain $\widetilde{\lambda}_{\Sigma}^2 (\Lambda_{G}) < 10^{-2}$
at the GUT scale.
We observe that the coupling $\lambda_{\Sigma}$ 
is not directly coupled to $\lambda_t$ and $\lambda_b$ and is
only weakly coupled to $\lambda_H$; 
so it is interesting to analyze the perturbation
on the exact fixed point predictions coming from a small
deviation along the direction $\lambda_{\Sigma}=0$.
The fixed point for the truncated system $(\lambda_t,\lambda_b,\lambda_H)$,
assuming that $\lambda_{\Sigma}$ is a constant parameter,
can be calculated from Eq.~(\ref{su5Yukgaurge}), and we obtain
%
\begin{equation}
\frac{\lambda_t^2}{g^2}  =   \frac{2533}{2605} + 
\frac{126}{2605}\frac{\lambda_{\Sigma}^2}{g^2},
\qquad 
\frac{\lambda_b^2}{g^2}  =   \frac{1491}{2605} + 
\frac{126}{2605}\frac{\lambda_{\Sigma}^2}{g^2},
\qquad
\frac{\lambda^2_{H}}{g^2}  =   \frac{560}{521} -
\frac{273}{2084}\frac{\lambda_{\Sigma}^2}{g^2}.
\label{su5fpyukpert}
\end{equation}
\begin{figure}
\begin{center}
\includegraphics[angle=0, width=0.8\textwidth]{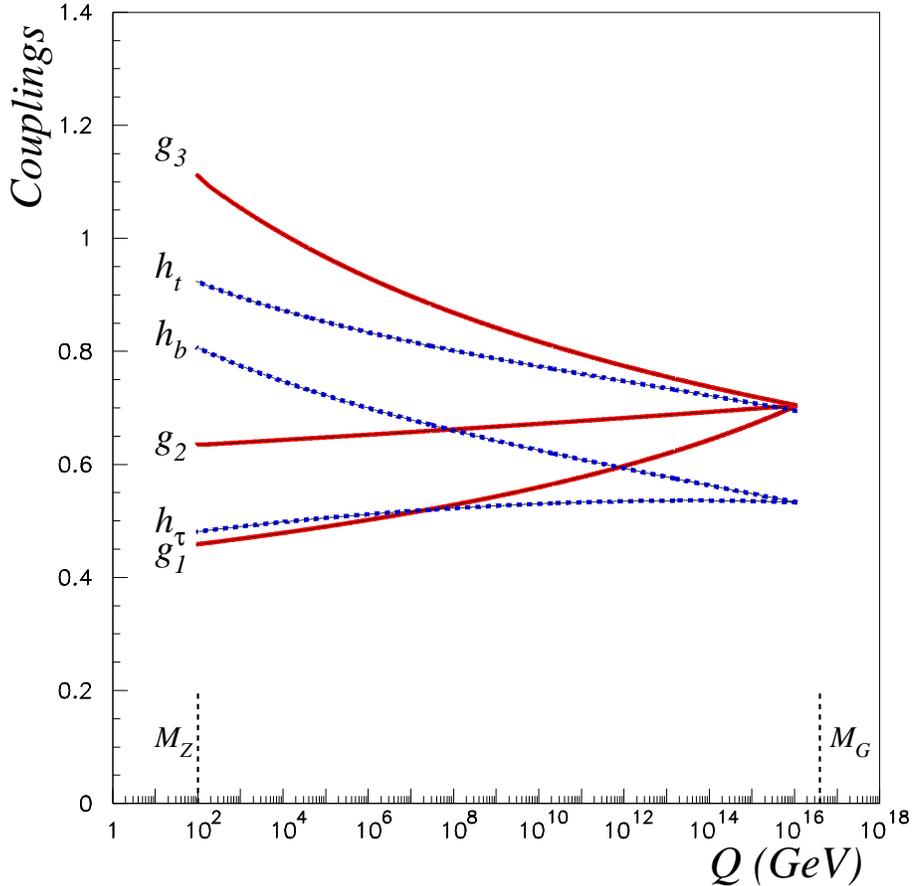}
\caption{\rm  Two-loop
gauge and Yukawa coupling evolution
from the unification scale to $M_Z$ scale
for the exact $SU(5)$ fixed point boundary conditions. 
The plot corresponds to 
point 1 in Table~\ref{su5pointsmun}.}
\label{fig:su5fpyuk}      
\end{center}
\end{figure}      
%
From these equations we see that the na\" \i ve phenomenological constraint
on proton decay, $\lambda_H>g$, implies an
upper bound on $\lambda_{\Sigma}^2$ given by 
$\lambda_{\Sigma}^2 < 4/7 ~g^2 = 0.5714 ~g^2$.
Therefore, for the fixed point to become phenomenologically 
viable, the coupling
$\lambda_{\Sigma}$ must be much smaller than $g$; 
otherwise we cannot overcome the constraints on proton decay. 
If $\lambda_{\Sigma}^2$ is less than $0.5~g^2$, 
the fixed point predictions for 
$\lambda_t$ and $\lambda_b$ at the GUT scale, 
given in Eqs.~(\ref{su5fpyukpert}), 
are constrained to be less than $2\%$ away 
from the exact fixed point predictions.
This perturbation, when extrapolated to the electroweak scale,
affects slightly the predictions for fermion masses, which
turn out to be less than $1\%$ away from the exact fixed point predictions.
Therefore,
small perturbations in the $\lambda_{\Sigma}$ direction
do not affect seriously the exact fixed point predictions, 
since the fermion Yukawa couplings 
are not directly coupled to $\lambda_{\Sigma}$.
From now on we will assume that at the GUT scale
$\lambda_{\Sigma}$ is much smaller than the other couplings:
\begin{equation}
\left| \lambda_{\Sigma} \right| << \lambda_t, \lambda_b, \lambda_H, g.
\label{pertassum}
\end{equation}
In this case it makes sense to study 
the predictions ensuing from the exact 
fixed point given by Eq.~(\ref{su5fpyuksol}),
even when $\lambda_{\Sigma}=0$ is a non attractive direction.

\subsection{Numerical results without supersymmetric thresholds}
We now turn to an analysis of 
the fixed point predictions for
the third generation fermion masses and the standard model gauge couplings.
We will assume that the SU(5) symmetry
breaks to the standard model group, $G_{SM}$, at the GUT scale
and, temporarily, 
ignore the effects of the low-energy 
supersymmetric thresholds 
to gauge and Yukawa coupling predictions. 
We assume that $\lambda_{\Sigma} << g$,
so that our model has only three basic parameters: the unification scale $M_G$,
the unified gauge coupling $g$, and 
the ratio of vacuum expectation values in the MSSM, $\tan\beta$.
$M_G$ and $g$ are high energy model parameters while $\tan\beta$ 
is a low-energy parameter. 
Below $M_G$,  the effective theory is the third generation MSSM
given by the superpotential
\begin{equation}
{\cal W}_{\rm MSSM} = \mu ~ {\cal H}_{u} {\cal H}_d
+  h_t {\cal Q} {\cal H}_u {\cal U} + h_b {\cal Q} {\cal H}_d {\cal D}
+ h_{\tau} {\cal L} {\cal H}_d {\cal E},
\label{mssmsup}
\end{equation}
where the Yukawa couplings at the GUT scale 
are determined by the boundary conditions
given by the SU(5) fixed point, Eq.~(\ref{su5fpyuksol}).
We will scan the parameter space and will require that
the strong gauge coupling, the fine structure constant and
the tau mass be in the measured range \cite{Groom:in}:
\begin{eqnarray} 
\widehat{\alpha}_e(m_Z)^{-1} &=& 127.934 \pm 0.027,  \\ 
\alpha_s(m_Z) &=&  0.1172 \pm 0.0020,  \\ 
m_{\tau} &=& 1.77703 ~{\rm GeV}.
\end{eqnarray}
\begin{table} \centering    
\begin{tabular}{|c|r|}
\hline \hline
\multicolumn{2}{|c|}{model parameters} \\ \hline
$M_{G}\times 10^{-16} $~(GeV) & $1.81 \pm 0.20$ \\
$g_G$ & $0.725 \pm 0.001$ \\ 
$\tan \beta$ & $47.98 \pm 0.03$ \\ \hline
\multicolumn{2}{|c|}{experimental constraints} \\ \hline
$\widehat{\alpha}_e(m_Z)^{-1}_{\overline{\rm MS}}$ & $127.934\pm 0.027$  \\ 
$\alpha_s(m_Z)_{\overline{\rm MS}}$ &  $0.1172 \pm 0.0020$  \\ 
$m_{\tau}$~(GeV) & $1.77703$  \\ \hline
\multicolumn{2}{|c|}{theoretical predictions} \\ \hline
$\widehat{s}_W^2(m_Z)_{\overline{\rm MS}}$ & $0.2334 \pm 0.0005$ \\
$m_t$~(GeV) & $181.5 \pm 1.2$ \\ 
$m_b(m_Z)_{\overline{\rm MS}}$~(GeV) & $3.20 \pm 0.02$ \\
\hline \hline
\end{tabular} 
  \caption{\rm Exact SU(5) fixed point predictions without including 
superspectrum thresholds. We fit the central values of 
$\widehat{\alpha}_e(m_Z)^{-1}_{\overline{\rm MS}}$, 
$\alpha_s(m_Z)_{\overline{\rm MS}}$ and $m_{\tau}$ with
experimental uncertainties and obtain three theoretical predictions
for $\widehat{s}_W^2(m_Z)$, $m_t$ and $m_b(m_Z)$, which must be compared
with the experimental measurements $m_t = 174.3 \pm 5.1$~GeV,
$m_b(m_Z^>)_{\overline{\rm MS}}= 2.86 \pm 0.2$~GeV and
$\widehat{s}_W^2(m_Z)_{\overline{\rm MS}} = 0.23117(16)$.}
  \label{su5tablenoth}    
\end{table} 
{\small
\begin{table} \centering    
\begin{tabular}{|c|c|}
\hline \hline
\multicolumn{2}{|c|}{model parameters} \\ 
\hline
$M_{G}\times 10^{-16}$~({\rm GeV}) &  1.8139  \\
$g$ &  0.7255  \\
$\tan \beta$ &  47.9864    \\ 
\hline 
\multicolumn{2}{|c|}{{\rm MSSM Yukawa couplings at $M_G$ (fixed point values)}} \\
\hline
$h_t(M_G)$ & 0.71548 \\
$h_b(M_G)$ & 0.54893  \\
$h_{\tau}(M_G)$ & 0.54893  \\
\hline
\multicolumn{2}{|c|}{{\rm MSSM dimensionless couplings at $M_Z$}} \\
\hline
$g_1(M_Z)$ & 0.46213  \\
$g_2(M_Z)$ & 0.64873  \\
$g_3(M_Z)$ & 1.21943  \\
$h_t(M_Z)$ & 0.98845  \\
$h_b(M_Z)$ & 0.87367  \\
$h_{\tau}(M_Z)$ & 0.48392 \\
\hline
\multicolumn{2}{|c|}{{\rm experimental constraints}} \\
\hline
$\alpha_e(m_Z)^{-1}_{\overline{\rm MS}}$ & 127.925 \\ 
$\alpha_s(m_Z)_{\overline{\rm MS}}$ &  0.11721 \\ 
$m_{\tau}$~({\rm GeV}) & 1.77703  \\ \hline
\multicolumn{2}{|c|}{{\rm theoretical predictions}} \\
\hline
$s_W^2(m_Z)_{\overline{\rm MS}}$ & 0.23340  \\
$m_t$~({\rm GeV}) & 181.631  \\ 
$m_b(m_Z)_{\overline{\rm MS}}$~({\rm GeV}) & 3.202  \\
\hline \hline
\end{tabular} 
  \caption{\rm Representative point of the fit for the exact SU(5) fixed point 
without supersymmetric thresholds.}
  \label{su5pointnoth}    
\end{table} 
}
Roughly speaking, the strong gauge coupling and the
fine structure constant determine the values of the
unification scale and the unified gauge coupling.
Once we know $M_G$ and $g$, the fixed point predicts
the value of the tau Yukawa coupling at the $m_Z$ scale.
We can then determine $\tan\beta$ 
using the tau mass measurement and the tree level
tau mass formula in the MSSM,
\begin{equation}
m_{\tau}(m_Z^>) = h_{\tau}(m_Z) v(m_Z) \cos\beta.
\label{taueq}
\end{equation}
Here, $v(m_Z)=2m_W/g_2(m_Z)$, $g_2(m_Z)$ is the 
weak coupling constant at $m_Z$, and $m_{\tau}(m_Z^>)$
is the tau running mass at $m_Z$ above the electroweak threshold.
The tau pole mass, $m_{\tau}$, 
must be related to the running mass below the electroweak threshold
, $m_{\tau}(m_Z^<)$, including QED corrections \cite{Arason:1991ic}
\begin{equation}
m_{\tau}(m_Z^<) = m_{\tau}\left\{ 1 - \frac{\alpha_e(m_Z)}{\pi}
\left(1 +\frac{3}{4}\ln \left( \frac{m_Z^2}{m_{\tau}(m_Z^<)} \right)
\right] \right\}.
\end{equation}
We must include the effects of $W$ and $Z$ bosons to obtain
the running mass just above the electroweak threshold,
\begin{equation}
m_{\tau}(m_Z^>) = m_{\tau}(m_Z^<) \left( 1 + \Delta^{\tau}_{SM}\right),
\end{equation}
where the expression for the electroweak thresholds, $\Delta^{\tau}_{SM}$, 
is given in Ref. \cite{Wright:1994qb} and is very small, approximately 
$0.1\%$. 
Using the particle data group (PDG) 
central values we find that the tau running mass 
is $m_{\tau}(m_Z^<)= 1.7463$~GeV just below $m_Z$
and $m_{\tau}(m_Z^>)= 1.7482$~GeV just above $m_Z$. 
Once we know $\tan\beta$, we obtain, 
along with the weak mixing angle,
two additional predictions from the SU(5) fixed point:
the top and bottom quark masses. The running bottom quark
mass just above $m_Z$ in the dimensional reduction scheme 
($\overline{\rm DR}$) is computed using
the tree level MSSM formula for the bottom mass,
\begin{equation}
m_{b}^{\overline{\rm DR}}(m_Z^>) = h_{b}(m_Z) v(m_Z) \cos\beta.
\end{equation}
To convert $m_b^{\overline{\rm DR}}(m_Z^>)$ to the minimal substraction
scheme ($\overline{\rm MS}$) we use the expression \cite{Baer:2002ek}
\begin{equation}
m_{b}^{\overline{\rm DR}}(m_Z^>) =  m_{b}^{\overline{\rm MS}}(m_Z^>)
\left[ 1 - \frac{1}{3} \frac{\alpha_s^{\overline{\rm MS}}(m_Z)}{\pi} 
- \frac{29}{72} \left( \frac{\alpha_s^{\overline{\rm MS}}(m_Z)}{\pi} \right)^2 
\right].
\end{equation}
We note that to compute $\alpha_s^{\overline{\rm MS}}(m_Z)$ we convert 
$\alpha_s^{\overline{\rm DR}}(m_Z)$ using \cite{Antoniadis:1982vr}
\begin{equation}
\frac{1}{\alpha_s^{\overline{\rm DR}}(m_Z)} =
\frac{1}{\alpha_s^{\overline{\rm MS}}(m_Z)}
- \frac{1}{4 \pi}.
\end{equation}
To obtain 
the experimental bottom running mass just above $m_Z$,
$m_{b}^{\overline{\rm MS}}(m_Z^>)$,
we start with the experimental value for 
$m_{b}^{\overline{\rm MS}}(m_b^{\overline{\rm MS}})$;
using the analytical solution of the SM RGE for 
the running mass we compute
the bottom running mass just below the electroweak threshold. Then
we include the electroweak loops, $\Delta^{b}_{SM}$, 
to obtain the running mass just above $m_Z$, 
\begin{equation}
m_{b}^{\overline{\rm MS}}(m_Z^>) = m_{b}^{\overline{\rm MS}}(m_Z^<) 
\left( 1 + \Delta^{b}_{SM}\right).
\end{equation}
Once again, the expression for the electroweak thresholds, $\Delta^{b}_{SM}$
is given in \cite{Wright:1994qb} and is approximately $-0.4\%$.
Using $m_{b}^{\overline{\rm MS}}(m_b^{\overline{\rm MS}})=4.2$~GeV
and central PDG values for gauge couplings, we obtain
$m_{b}^{\overline{\rm MS}}(m_Z^<)=2.87$~GeV just below $m_Z$, and 
$m_{b}^{\overline{\rm MS}}(m_Z^>)=2.86$~GeV just above $m_Z$, 
and after $\overline{\rm MS}$ to $\overline{\rm DR}$ conversion 
$m_{b}^{\overline{\rm DR}}(m_Z^>)=2.82$~GeV. This is the
value to which we should compare our prediction.
The physical top quark mass is predicted using the formula
\begin{equation}
m_t = h_t(m_t) v(m_t) \sin\beta \left( 1 + \Delta_{\rm QCD}^{\rm t} \right),
\end{equation}
where $\Delta_{\rm QCD}^{\rm t}$ stands for the gluon correction to the
top mass evaluated at the $m_t$ scale.
In the $\overline{\rm DR}$ scheme this is given by
\begin{equation}
\Delta_{\rm QCD}^{\rm t} = \frac{5}{3} \frac{\alpha_s(m_t)}{\pi} 
+ 8.15 \left(\frac{\alpha_s(m_t)}{\pi}\right)^2
+ 71.5 \left(\frac{\alpha_s(m_t)}{\pi}\right)^3.
\label{gluoncor}
\end{equation}
We observe that the order $\alpha_s$ coefficient is $5/3$
in the $\overline{\rm DR}$ scheme and
$4/3$ in the $\overline{\rm MS}$ scheme \cite{Tarrach:1980up}. 
The order $\alpha_s^2$ coefficient in the $\overline{\rm DR}$ 
scheme has been computed
from Ref.~\cite{Avdeev:1997sz}. The order $\alpha_s^3$ term is
the coefficient computed in the $\overline{\rm MS}$ scheme and
can be extracted from Ref.~\cite{Melnikov:2000qh}.
The difference between $v(m_t)$ and $v(M_Z)$ is only of the order
$$
v(m_Z)-v(m_t) \simeq \frac{2}{3}\ln\left( \frac{m_t}{m_Z} \right)~\hbox{\rm GeV},
$$
which represents around $0.25$\%. 
In practice, we implement this procedure numerically using
two-loop MSSM RGEs for gauge and Yukawa couplings 
from the unification to the $m_Z$ scale \cite{rges,isasugra}.
Our numerical results can be read off Table~\ref{su5tablenoth}.
The uncertainties in the model parameters and theoretical predictions
shown in Table~\ref{su5tablenoth}
are theoretical errors induced by uncertainties in
the experimental constraints.
Our fixed point predictions, neglecting supersymmetric thresholds,
thus become 
\begin{eqnarray} 
m_t &=& 181.6 \pm 1.3 ~{\rm GeV}, \\ 
m_b(m_Z)_{\overline{\rm MS}} &=& 3.20 \pm 0.03~{\rm GeV}, \\  
\widehat{s}_W^2(m_Z)_{\overline{\rm MS}} &=& 0.2334 \pm 0.0005.
\end{eqnarray}
These must be compared with the
experimental measurements \cite{Groom:in}
\begin{eqnarray} 
m_t &=& 174.3 \pm 5.1 ~{\rm GeV}, \\ 
m_b(m_Z^>)_{\overline{\rm MS}} &=& 2.86 \pm 0.2~{\rm GeV}, \\  
\widehat{s}_W^2(m_Z)_{\overline{\rm MS}} &=& 0.23117(16).
\end{eqnarray}
Surprisingly, the tree level (i.e. with no SUSY thresholds)
fixed point prediction is 3~\% away from the
experimental central value for the top mass 
and 4~\% away from the experimental central value for
the bottom mass.
For comparision, we show in Table~\ref{su5pointnoth}
the high-energy and
low-energy coupling predictions for one of the
central points of the fit.
The relatively accurate prediction for the bottom mass at this stage is not
a surprise, since the bottom--tau mass ratio 
is a successful classical prediction of SUSY GUTs when SUSY threshold
corrections are not included \cite{bottomtau}.
Not taking into account these corrections, however, renders 
the classical prediction for the bottom-tau mass ratio irrelevant,
since the contribution of SUSY thresholds to the bottom mass can be 
very important \cite{thresholds} -- as much as  $\pm 40\%$ for large
$\tan\beta$, depending on the sign of the $\mu$ term. 
The case of the top mass prediction is similar, although 
we had no reason, a priori, to expect a good agreement with data. 
This good prediction must also be considered
only a provisional success, as
we have ignored the effect of
supersymmetric thresholds.
For the moment,
we do not know if the inclusion of these 
thresholds would spoil 
our interesting predictions.
A predictive theory for the supersymmetry breaking sector
is necessary in order to 
include these supersymmetric thresholds.
%
\section{minimal SU(5) fixed point and soft breaking\label{sec3}}
%
A realistic globally supersymmetric 
SU(5) model can be constructed \cite{Dimopoulos:1981zb},
in which 
supersymmetry is broken explicitly, but softly, by
terms of dimension less than four in the Lagrangian 
\cite{Girardello:1981wz}.
The scalar potential will contain the following
SU(5)-invariant terms:
\begin{eqnarray}
{\cal V_{\rm SU(5)}} &=& m^2_{H_{\overline{5}}} \left| {h}_{\overline{5}} \right|^2 +
 m^2_{H_5}\left| {h}_5 \right|^2 +
 m^2_{\Sigma}~{\rm tr} \left\{ \sigma^{\dagger}\sigma \right\}+
 m^2_{5}\left| \phi_{\overline{5}} \right|^2 +
 m^2_{10}~{\rm tr} \left\{ \phi^{\dagger}_{10}\phi_{10} \right\} 
\nonumber  \\
&+&\left[
\frac{1}{6}A_{\Sigma}\lambda_{\Sigma}~{\rm tr} \sigma^3 
+A_{H}\lambda_{H}~h_{\overline{5}}\sigma h_5
+ \frac{1}{4} A_t \lambda_t \phi_{10} \phi_{10}  {h}_5
+ \sqrt{2} A_b \lambda_b \phi_{10} \phi_{\overline{5}} {h}_{\overline{5}} \right.\\
&+&\left. 
\frac{1}{2} M_{1/2} \lambda_{\alpha}\lambda_{\alpha}
+B_{\Sigma}\mu_{\Sigma}~{\rm tr} \sigma^2 
+B_{H}\mu_{H} h_{5}h_{\overline{5}} + {\cal {\rm h.c.}}  \right], \nonumber
\end{eqnarray}
where $h_5$, $h_{\overline{5}}$, $\sigma$, $\phi_{\overline{5}}$ and
$\phi_{10}$ are the scalar components of the superfields
 ${\cal H}_u$, ${\cal H}_d$, $\Sigma$, $\psi_{\overline{5}}$ and
$\psi_{10}$, respectively, and $\lambda_{\alpha}$ are the SU(5) 
gaugino fields.
It is known that 
the purpose of mass terms 
for matter and gaugino fields 
is to increase the mass of the still-unobserved MSSM spectra at low-energy. 
The soft sector parameters must be chosen with care
to assure that the Hamiltonian is bounded from below and to obtain a 
desirable pattern of SU(5) breaking
\cite{susysu5breaking}.
In principle the soft parameters are arbitrary.
Interestingly, however, it has been pointed out that the existence 
of infrared stable
fixed points for the Yukawa couplings implies (given asymptotic freedom)
infrared stable fixed points for the soft parameters \cite{Jack:1998xw}.
Moreover, at this fixed point the dimensionful 
soft-breaking parameters are all determined
by the gaugino mass. Special relationships then 
appear, and as a consequence a particular pattern of supersymmetric
spectra emerges at low-energy if electroweak symmetry breaking can
be achieved \cite{jackjones}.
We begin by analyzing the one-loop RGEs for the soft SUSY breaking parameters,
with the goal of identifying their infrared fixed point
associated with the fixed point of the Yukawa couplings.
The fixed points for the trilinear soft parameters
will appear for the ratios of trilinear parameters to 
the gaugino unified mass \cite{Lanzagorta:1995ai}.
The RGEs for these ratios are
\begin{equation}
\frac{1}{g^2}\frac{d}{dt}
\left(
\begin{array}{c}
\widetilde{A}_t \\
\widetilde{A}_b \\
\widetilde{A}_{H} \\
\widetilde{A}_{\Sigma} 
\end{array}
\right) = 
\left(
\left(
\begin{array}{cccc}
9 \widetilde{\lambda}_t^2 -b 
& 4 \widetilde{\lambda}_b^2 & 24/5  \widetilde{\lambda}^2_H & 0 \\
3 \widetilde{\lambda}_t^2 & 10 \widetilde{\lambda}_b^2 -b 
& 24/5 \widetilde{\lambda}^2_H & 0 \\
3 \widetilde{\lambda}_t^2 & 4 \widetilde{\lambda}_b^2 & 
53/5 \widetilde{\lambda}^2_H -b
& 21/20 \widetilde{\lambda}^{2}_{\Sigma} \\
0 & 0 & 3 \widetilde{\lambda}^2_H & 
63/20 \widetilde{\lambda}^{2}_{\Sigma} -b
\end{array}
\right)
\left(
\begin{array}{c}
\widetilde{A}_t \\
\widetilde{A}_b \\
\widetilde{A}_{H} \\
\widetilde{A}_{\Sigma} 
\end{array}
\right)
+
\left(
\begin{array}{c}
96/5 \\
84/5 \\
98/5 \\
30 
\end{array}
\right)
\right),
\label{su5trilrge}
\end{equation}
%
where we have defined
%
\begin{equation}
\widetilde{A}_t =
\frac{A_t}{M_{1/2}},
\qquad
\widetilde{A}_b =
\frac{A_b}{M_{1/2}},
\qquad
\widetilde{A}_{H} =
\frac{A_{H}}{M_{1/2}},
\qquad
\widetilde{A}_{\Sigma} =
\frac{A_{\Sigma}}{M_{1/2}}.
\label{trilrat}
\end{equation}
We observe again, from Eq.~(\ref{su5trilrge}),
that the $\lambda_{\Sigma}$ is not
coupled directly to $A_t$ and $A_b$ and is only weakly coupled to
$A_{H}$. 
There will be a
fixed point for the soft trilinears associated with the
fixed point, Eq.~(\ref{su5fpyuksol}), in the Yukawa sector 
given by
%
\begin{equation}
\frac{A_t}{M_{1/2}}  =  -1,
\qquad
\frac{A_b}{M_{1/2}}  =  -1,
\qquad
\frac{A_{H}}{M_{1/2}}  =  -1,
\qquad
\frac{A_{\Sigma}}{M_{1/2}}  = -\frac{4650}{521}.
\label{su5fptrilsol}
\end{equation}
%
\begin{figure}
\begin{center}
\includegraphics[angle=0, width=0.8\textwidth]{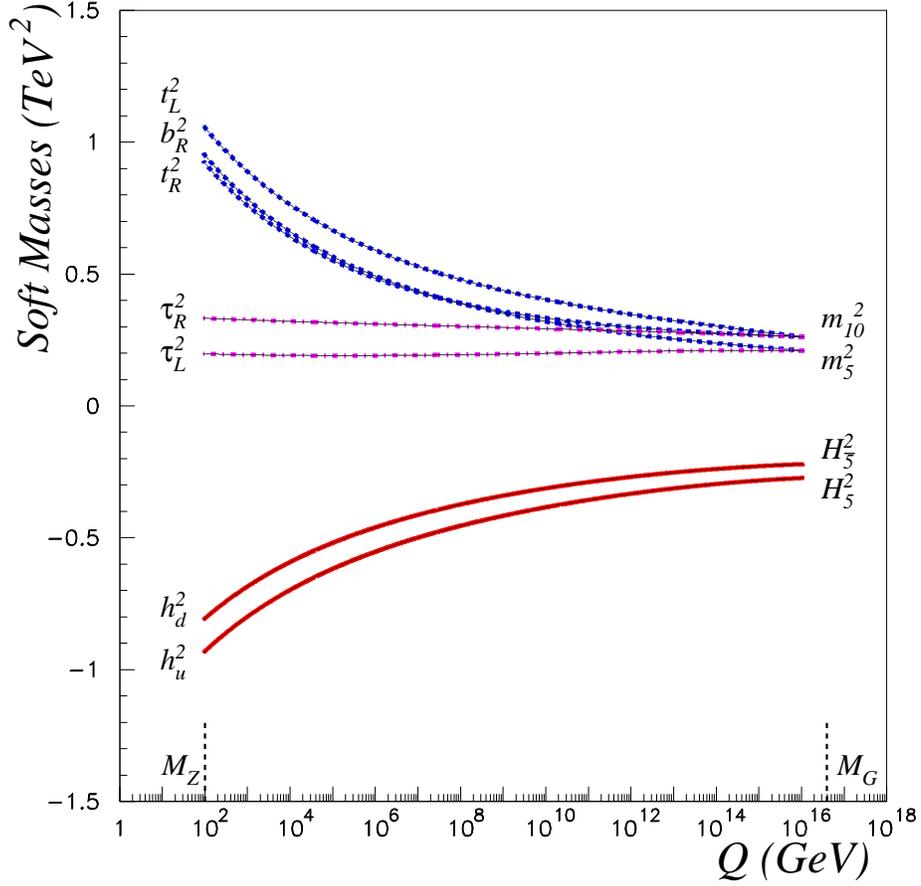}
\caption{\rm The two-loop evolution from unification to $M_Z$ scale of 
soft masses for the exact $SU(5)$ fixed point. The 
plot corresponds to point 1 in Table~\ref{su5pointsmun}.}
\label{fig:su5fpsoft}      
\end{center}
\end{figure}      
%
Analogously, fixed points for the soft masses
will appear for the squared ratios of soft masses to 
the gaugino unified mass \cite{Lanzagorta:1995ai}.
The RGEs for these ratios in the minimal SU(5) model are given by
\begin{equation}
\frac{1}{g^2}\frac{d}{dt}
\left(
\begin{array}{c}
\widetilde{m}_5^2 \\
\widetilde{m}_{10}^2 \\
\widetilde{m}_{{\cal H}_5}^2 \\
\widetilde{m}_{{\cal H}_{\overline{5}}}^2 \\
\widetilde{m}_{\Sigma}^2 
\end{array}
\right) = 
\left(
{\cal M}
\left(
\begin{array}{c}
\widetilde{m}_5^2 \\
\widetilde{m}_{10}^2 \\
\widetilde{m}_{{\cal H}_5}^2 \\
\widetilde{m}_{{\cal H}_{\overline{5}}}^2 \\
\widetilde{m}_{\Sigma}^2 
\end{array}
\right)
+
{\cal N}
\left(
\begin{array}{c}
\widetilde{A}_t^2 \\
\widetilde{A}_b^2 \\
\widetilde{A}_{H}^2 \\
\widetilde{A}_{\Sigma}^2 
\end{array}
\right)
-
\left(
\begin{array}{c}
48/5 \\
72/5 \\
48/5 \\
48/5 \\
20 
\end{array}
\right)
\right),
\label{su5softmasrge}
\end{equation}
%
where we have defined
%
\begin{equation}
\widetilde{m}_5^2 =
\frac{m_5^2}{M^2_{1/2}},
\qquad
\widetilde{m}_{10}^2 =
\frac{m_{10}^2}{M^2_{1/2}},
\qquad
\widetilde{m}_{{\cal H}_5}^2 =
\frac{m_{{\cal H}_5}^2}{M^2_{1/2}},
\qquad
\widetilde{m}_{{\cal H}_{\overline{5}}}^2 = 
\frac{m_{{\cal H}_{\overline{5}}}^2}{M^2_{1/2}},
\qquad
\widetilde{m}_{\Sigma}^2 =
\frac{m_{\Sigma}^2}{M^2_{1/2}},
\label{sfmasrat}
\end{equation}
and
\begin{equation}
{\cal N} =
\left(
\begin{array}{cccc}
0 & 4 \widetilde{\lambda}_b^2 & 0 & 0 \\
3 \widetilde{\lambda}_t^2 & 2 \widetilde{\lambda}_b^2 & 0 & 0 \\
3 \widetilde{\lambda}_t^2 & 0 & 24\widetilde{\lambda}_H^2/5 & 0 \\
0 & 4 \widetilde{\lambda}_b^2 & 24\widetilde{\lambda}_H^2/5 & 0 \\
0 & 0 & \widetilde{\lambda}^2_H &  21 \widetilde{\lambda}^{2}_{\Sigma}/20
\end{array}
\right),
\label{softmatM}
\end{equation}
\begin{equation}
{\cal M} =
\left(
\begin{array}{ccccc}
4 \widetilde{\lambda}_b^2 
& 4 \widetilde{\lambda}_b^2 & 0 & 4\widetilde{\lambda}_b^2 & 0 \\
2 \widetilde{\lambda}_b^2 & 6 \widetilde{\lambda}_t^2 +
2 \widetilde{\lambda}_b^2  
& 3 \widetilde{\lambda}_t^2 & 2\widetilde{\lambda}_b^2 & 0 \\
0 & 6 \widetilde{\lambda}_t^2 & 3 \widetilde{\lambda}_t^2 +
24\widetilde{\lambda}_H^2/5  & 
24 \widetilde{\lambda}_H^2/5
& 24\widetilde{\lambda}_H^{ 2}/5 \\
4 \widetilde{\lambda}_b^2 & 4 \widetilde{\lambda}_b^2 & 
24 \widetilde{\lambda}_H^{2}/5
& 4 \widetilde{\lambda}_b^2 + 24 \widetilde{\lambda}_H^2/5 & 
24 \widetilde{\lambda}_H^2 /5 \\
0 & 0 & \widetilde{\lambda}_H^2 &  \widetilde{\lambda}_H^2 &
\widetilde{\lambda}_H^2 + 63 \widetilde{\lambda}_{\Sigma}^{2}/20 
\end{array}
\right) - 2b ~{\cal I}_5,
\label{softmatN}
\end{equation}
where ${\cal I}_5$ is the $5\times 5$ identity matrix.
From Eq.~(\ref{su5softmasrge}) it is trivial to calculate
the fixed point solution associated with the 
fixed point, Eq.~(\ref{su5fpyuksol}),
in the Yukawa sector:
%
\begin{equation}
\frac{m_5^2}{M^2_{1/2}} = \frac{436}{521}, 
\quad
\frac{m_{10}^2}{M^2_{1/2}} =  \frac{545}{521} ,
\quad
\frac{m_{{\cal H}_5}^2}{M^2_{1/2}}= -\frac{569}{521} ,
\quad
\frac{m_{{\cal H}_{\overline{5}}}^2}{M^2_{1/2}} = -\frac{460}{521} ,
\quad
\frac{m_{\Sigma}^2}{M^2_{1/2}} =  \frac{1550}{521}.
\label{su5fpsoftmassol}
\end{equation}
It can be easily proven that if the Yukawa couplings 
evolve towards the fixed point solution given by Eq.~(\ref{su5fpyuksol}),
the associated minimal SU(5) soft fixed point
given by Eqs.~(\ref{su5fptrilsol}) and (\ref{su5fpsoftmassol})
is stable~\cite{Jack:1998xw}.
We already observed that the Yukawa fixed point, Eq.~\ref{su5fpyuksol},
in the direction $\lambda_{\Sigma}=0$ is not attractive.
It is therefore crucial for the consistency of the soft-sector 
fixed point predictions  
that $\lambda_{\Sigma}$ be much smaller than
the other dimensionless couplings. 
We also observe that
the parameters
$\mu_H$, $\mu_{\Sigma}$, $B_H$, and $B_{\Sigma}$ 
decouple from the rest of RGEs. 
To sum up, if we assume that the soft sector 
is at the fixed point associated
with the fixed point in the Yukawa sector, the complete soft sector
turns out to be strongly constrained. 
Therefore, we expect that definite predictions 
for the third generation supersymmetric spectra,
as functions of the gaugino unified mass,
will arise from the minimal SU(5) fixed point.
In Fig.~\ref{fig:su5fpsoft} we show the evolution 
of soft masses predicted by the fixed point 
for a representative value with $M_{1/2}=500$~GeV.

Let us add some qualitative comments regarding the uncertainty arising from the
unknown ranges for the input conditions at a higher scale (Planck scale).
When there is one dominant Yukawa coupling,
one can address this question qualitatively.
The rate of approach to the fixed point in this case is related to  
the coefficient $\left( \alpha_s(0) / \alpha_s(M_P) \right)^B$, with $B= 1 + (r/b)$, 
where $b$ is the $\alpha_s$ beta coefficient and $r$ is the $\alpha_s$ coefficient
in the Yukawa coupling RGE. The smaller this coefficient is, the faster the approach
to the infrared fixed point is.  In the minimal SUSY SU(5), $b=-3$ and $r = 96/5$ and $84/5$
for the top and bottom-tau Yukawa couplings respectively. 
We can compare this case with the low $\tan(\beta)$ MSSM fixed point,
where $r=16/3$ for the top Yukawa coupling. 
Since $\left( \alpha_s(0) / \alpha_s(M_P) \right) >1$ 
and $B<0$ and approximately $8$
times larger in the minimal SUSY SU(5), we see that the coefficient 
is smaller in the minimal SUSY SU(5) than in the MSSM low tan(beta)
fixed point scenario, i.e. the approach to the fixed point is faster 
in the minimal SUSY SU(5).
Furthermore, we must keep in mind that 
in the case where several Yukawa couplings are of the same order,
analytical approximations may not be reliable; a lengthy numerical study
is required to determine the vessel attraction of the fixed point. 
We note, in favor of our main hypothesis, that the SU(5) fixed point predicts approximate universality
for the sfermion soft masses, approximate top--bottom--tau 
Yukawa unification, and negative soft Higgs boson masses squared.
It should be pointed out that many SUGRA models predict universality for
the soft masses close to the Planck scale, and some string--inspired SUGRA models predict
Yukawa unification. Observe that if we assume we have a SUGRA model at Planck scale
with soft mass universality and top--bottom--tau Yukawa unification, then
the initial conditions at Planck scale would already be very close to
the SU(5) fixed point patterns, and the convergence to the fixed point at GUT scale could be much faster.
With regard to the soft squared Higgs boson masses, even if these are positive
at the Planck scale, they can be driven quickly to negative values at the 
GUT scale. This was observed in the analysis of the nonuniversalities in the SU(5)
soft terms implemented by Polonsky and Pomarol (see Fig. 1 of Ref.~\cite{pomarol}).

\section{Unified fixed point and SU(5) breaking\label{sec4}}
%
The spontaneous symmetry breaking pattern is one of the non predictive
aspects of grand unified models \cite{su5breaking}. 
For the minimal supersymmetric SU(5), 
independent of the parameters
of the scalar potential, there are many vacuum expectation values,
all of which leave the supersymmetry unbroken. These
break SU(5) spontaneously to smaller groups: SU(4)$\times$U(1),
SU(3)$\times$SU(2)$\times$U(1), etc.
A vacuum expectation value (VEV) with the desired breaking pattern
to  SU(3)$\times$SU(2)$\times$U(1) is among the possible
vacua of the supersymmetric model
\begin{equation}
\left< \sigma \right> = \frac{\mu_{\Sigma}}{\lambda_{\Sigma}} \hbox{diag}
\left(2,2,2,-3,-3\right).
\label{su5vev}
\end{equation}
%
This interesting VEV is degenerate in energy
with a long list of physically unacceptable VEVs.
To remove the degeneracy and to break supersymmetry, 
soft terms are added to the scalar potential. 
These terms have to be chosen with care to obtain a
potential that remains bounded from below and leads
to a VEV with the desired SU(5) breaking pattern.
Certain relations between the scalar coupling constants of
the Higgs multiplets have to be satisfied in order
to have the SU(3)$\times$SU(2)$\times$U(1) minimum
as the absolute minimum of the potential.
In general, $m_{\Sigma}^2$ must be positive and
small (relative to the GUT scale) 
for phenomenologically interesting symmetry breaking.
One is thus led to take positive $m_5^2$ and $m_{10}^2$ to disfavor
energetically all the solutions that involve nonvanishing VEVs
for $\phi_{\overline{5}}$ and $\phi_{10}$ \cite{susysu5breaking}. 
These two necessary conditions are satisfied at the SU(5) fixed point 
given by Eqs.~(\ref{su5fpsoftmassol}).

It can be argued that some spontaneous symmetry breaking directions
could be more natural than others if an infrared attractive
fixed point is found within the corresponding region of Higgs
parameter space \cite{Amelino-Camelia:1996ax,amelino}.
The direction of SU(5) breaking determined by the vacuum 
expectation value $\left< \sigma \right>$ can be read off
the scalar potential at the minimum
\begin{equation}
{\cal V}^{{\rm min}}_{\rm SU(5)} = \rho M^2_{1/2} \left< \sigma \right>^2 \left(
\frac{m^2_{\Sigma}}{M^2_{1/2}} - \left[ \frac{1}{3}\frac{A_{\Sigma}}{M_{1/2}}
-2 \frac{B_{\Sigma}}{M_{1/2}} \right]
\frac{\lambda_{\Sigma}\left< \sigma \right>}{M_{1/2}}
\right),
\end{equation}
where $\rho=0$ for the minimum-preserving full SU(5) invariance,
$\rho=30$ for the $G_{SM}=$~SU(3)$\times$SU(2)$\times$U(1) invariant
minimum, and $\rho=20/9$ for the SU(4)$\times$U(1) invariant minimum.
If ${\cal V}<0$ the $G_{SM}$ invariant minimum is the lowest one,
while SU(5) remains unbroken if ${\cal V}>0$.
If $\left| \lambda_{\Sigma} \right| << 
\lambda_{t}, \lambda_{b}, \lambda_{H}, g$, the fixed point
predicts the ratios $m^2_{\Sigma}/M^2_{1/2}= 1550/521$, and 
$A_{\Sigma}/M_{1/2}= -4650/521$. 

The fixed point prediction
for $B_{\Sigma}/M_{1/2}$ can be calculated from the RGE
%
\begin{equation}
\frac{1}{g^2} \frac{d}{dt} \left( \frac{B_{\Sigma}}{M_{1/2}} \right)
= 4 \frac{A_{H}}{M_{1/2}}\frac{\lambda_{H}^2}{g^2} + 
\frac{84}{20} \frac{A_{\Sigma}}{M_{1/2}}\frac{\lambda_{\Sigma}^2}{g^2} + 40
+ b \frac{B_{\Sigma}}{M_{1/2}}.
\label{brge}
\end{equation}
%
At the fixed point, given by Eqs.~(\ref{su5fpyuksol}) and
\ref{su5fptrilsol}, 
$B_{\Sigma}/M_{1/2} = -6200/521= 4 A_{\Sigma}/3 M_{1/2}$. 
In addition, 
the SU(5) spectrum includes 12 
gauge bosons, $V=X,Y$, which receive a mass
$M_V= 5 g \left< \sigma \right> \gtrsim M_G $, while $M_{1/2}$
determines the masses of the supersymmetric spectra.
We therefore expect the ratio $ \left< \sigma \right>/M_{1/2}$ to be 
approximately $10^{12-14}$ by phenomenological constraints.
Therefore, the second or third
terms are individually many orders of magnitude larger than the first term.
As a consequence, the sign of the scalar potential is
sensitive to the sign of $\lambda_{\Sigma}$,
and we can only derive a constraint on $\lambda_{\Sigma}$
to obtain a satisfactory SU(5) breaking to the 
$G_{SM}$ invariant minimum
$$
{\cal V}^{{\rm min}}_{\rm SU(5)} 
< 0 \Longrightarrow \lambda_{\Sigma} > \frac{1}{3647}
\frac{M_{1/2}}{\left< \sigma \right>} \simeq 3 \times 10^{-16/-18}.
$$
Thus, a correct SU(5) breaking
is automatically achieved at the fixed point 
only for one sign of $\lambda_{\Sigma}$, 
positive in the convention used in this paper. 
We observe that the $\mu$-parameters, $\mu_H$ and $\mu_{\Sigma}$,
remain undetermined at the fixed point.

It is, however, well known that
in order for the Higgs SU(2) doublets to have masses
of ${\cal O}(M_Z)$ instead of ${\cal O}(M_G)$,
a fine tuning in the superpotential parameters is required,
$\mu = (\mu_H + 3 \lambda_H \left< \sigma \right> )\simeq {\cal O}(M_Z)$,
and one obtains $\mu_H \simeq - 3 \sqrt{560/521} ~g\left< \sigma \right>$
at the fixed point.
Additionally, a fine tuning condition also arises in the soft
sector, $B \mu = (B_H \mu_H + 3 A_H \lambda_H \left< \sigma \right> )
\simeq {\cal O}(M_Z^2)$, implying (using the first fine tuning
condition) $B_H = A_H$. This condition is slightly different from the
fixed point prediction for $B_H$, $B_H= 4A_H /3$, 
even though it implies the 
same sign constraint on $\lambda_{\Sigma}$; this indicates a possible
deviation from the fixed point prediction for the $B_H$ term.
To complete the GUT spectra, we mention that the
Higgs triplets receive a mass $M_{H_C}= 5\lambda_H \left< \sigma \right>$,
and the $\Sigma$ superfield decomposes with respect to
${\rm SU(3)}_C \times {\rm SU(2)}_L$ 
into $(3,2)+(\overline{3},2)+(8,1)+(1,3)+(1,1)$.
In the supersymmetric limit the $(3,2)$ and $(\overline{3},2)$
are degenerate with the gauge bosons; the $(1,3)$ and $(8,1)$
components, $\Sigma_3$ and $\Sigma_8$, respectively,
have a common mass $10\left| \mu_{\Sigma} \right|$;
and the mass of the singlet, $\Sigma_1$, is $2\left| \mu_{\Sigma} \right|$.
When soft breaking is considered, a boson-fermion mass splitting
is induced within every multiplet. 
%
\section{Electroweak breaking and supersymmetric spectra\label{sec5}}
%
\begin{figure}
\begin{center}
\includegraphics[angle=0, width=0.8\textwidth]{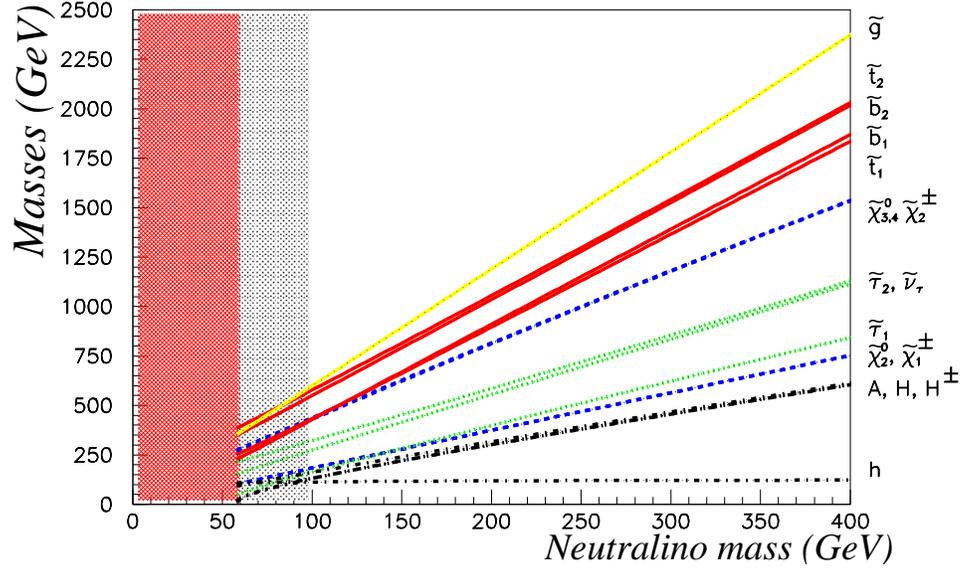}
\caption{\rm       
Low-energy superspectra from the exact SU(5) fixed point,
Eqs.~(\ref{boundarysoft1})-(\ref{boundarysoft4}). 
The dark shaded
region (red) is excluded by constraints on 
electroweak symmetry breaking, and the 
dot-shaded area (grey) is excluded by the experimental constraints
on the Higgs mass.}
\label{fig:su5spectra}      
\end{center}
\end{figure}      
%
\begin{figure}
\begin{center}
\includegraphics[angle=0, width=0.8\textwidth]{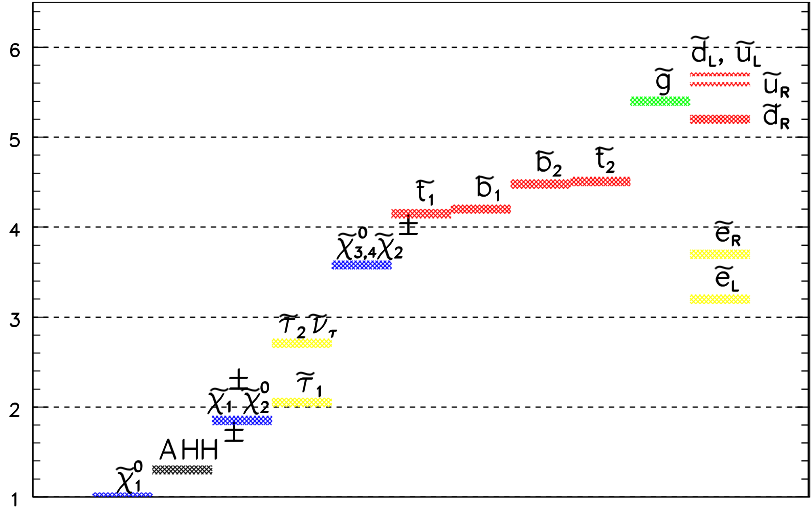}
\caption{\rm 
Characteristic ratios of masses for the low-energy superspectra from the 
exact SU(5) fixed point,
Eqs.~(\ref{boundarysoft1})-(\ref{boundarysoft4}). 
The superspectra 
for the first and second generation masses are computed 
assuming the trivial fixed point of the SU(5) RGEs
for the first and second generation soft parameters.}
\label{fig:ratsu5spectra}      
\end{center}
\end{figure}      
%
In order to analyze the implications of the SU(5) fixed point
in the supersymmetric spectra, we have to run the 
MSSM RGEs from the GUT to the $m_Z$ scale.
Below $M_G$  the effective theory is the third generation MSSM
defined by the superpotential given in Eq.~(\ref{mssmsup}).
The associated soft scalar potential is given by
\begin{eqnarray}
{\cal V}^{\rm soft}_{\rm MSSM} &=& m^2_{H_d}\left| {h}_d \right|^2 +
 m^2_{H_u}\left| {h}_u \right|^2 +
\frac{1}{2} \left[ \sum_{i=1,3} M_{i} \lambda_{i}\lambda_{i}
+B \mu h_{u}h_{d} + {\cal {\rm h.c.}}  \right]
\nonumber  \\
& &
+ m^2_{\widetilde{b}_L} \left| \widetilde{b}_L \right|^2 +
 m^2_{\widetilde{b}_R} \left| \widetilde{b}_R \right|^2 +
 m^2_{\widetilde{t}_L} \left| \widetilde{t}_L \right|^2 +
 m^2_{\widetilde{t}_R} \left| \widetilde{t}_R \right|^2 +
 m^2_{\widetilde{\tau}_R} \left| \widetilde{\tau}_R \right|^2 +
 m^2_{\widetilde{\tau}_L} \left| \widetilde{\tau}_L \right|^2
\nonumber  \\
& & +\left[ 
 A_t h_t \widetilde{t}_L {h}_u \widetilde{t}_R 
+ A_b h_b \widetilde{t}_L {h}_d \widetilde{b}_R 
+ A_{\tau} h_{\tau} \widetilde{\tau}_L {h}_d \widetilde{\tau}_R 
+ {\cal {\rm h.c.}} \right].
\end{eqnarray}
The soft parameters are determined at $M_G$
by the boundary conditions given by the SU(5) fixed point
at the unification scale, Eqs.~(\ref{su5fptrilsol})
and (\ref{su5fpsoftmassol})
%
\begin{eqnarray}
 m^2_{\widetilde{t}_L, \widetilde{t}_R, \widetilde{b}_L, \widetilde{\tau}_R}&=&
\frac{545}{521}~M^2_{1/2}, 
\label{boundarysoft1}
\\
m^2_{\widetilde{b}_R, \widetilde{\tau}_L, \widetilde{\nu}_{\tau}}&=&
\frac{436}{521}~M^2_{1/2}, 
\label{boundarysoft2}
\\
m_{{\cal H}_u}^2 = -\frac{569}{521}~M^2_{1/2} & <&
m_{{\cal H}_{d}}^2 = -\frac{460}{521}~M^2_{1/2}, 
\label{boundarysoft3}
\\
A_{t, b, \tau}  &=& - M_{1/2}.
\label{boundarysoft4}
\end{eqnarray}
The bilinear parameters $\mu$ and $B$ are not determined
by the fixed point. They must be computed from the MSSM 
minimization equations under the constraint  
of a correct electroweak symmetry breaking at $m_Z$.
In this scenario, as in generic supersymmetric unified scenarios,
we do not know why $m_Z$ is many orders of 
magnitude smaller than $M_G$; equivalently,
the MSSM $\mu$-term
must be many orders of magnitude smaller than $M_G$.
Many possible solutions have been 
proposed in the literature in the context of non minimal models
\cite{Kim:1983dt}. 

Let us describe in more detail the numerical procedure followed in this
paper to minimize the MSSM scalar potential. 
Once we assume some GUT scale initial conditions resulting from the SU(5) fixed point
we use the complete set of MSSM RGEs (two loop for dimensionless couplings
and main two loop contributions for dimensionful couplings) 
to extrapolate the MSSM couplings to the electroweak scale.
Next we use the experimental value of the tau mass and the tau Yukawa coupling at $m_Z$ 
(from the RGEs) to determine $\tan(\beta)$ at tree level using Eq.~(\ref{taueq}).
We use then the tree level minimization equations and the running couplings at $m_Z$,
obtained from the RGEs, 
to compute the parameters $\mu$ and $B\mu$ (this is the optimal choice since the
bilinear parameters are decoupled from the rest of RGEs), then we compute 
the tree level supersymmetric spectra.
Then we do a second iteration. Using the effective potential 
we implement the one--loop minimization of the scalar potential 
including corrections coming from sfermion loops \cite{Arnowitt:qp}, 
and we also include one loop SUSY threshold corrections in the
relation between third fermion masses and Yukawa couplings. 
Once we have included one loop corrections 
we compute again $\tan\beta$, $\mu$, $B\mu$ and the superspectra. 
We do several more iterations until
the results converge. In this way the correct
minimization of the MSSM, compatible with the large value of 
$\tan(\beta)$ fixed by the
tau mass, is guaranteed. 

We point out that the SU(5) fixed point boundary conditions
allow us to obtain a satisfactory tree-level electroweak symmetry breaking.
This is a non trivial fact that is quite simple to understand by a tree level analysis.
Even though the $\mu$-term and the $CP$-odd Higgs boson mass, $m_A$, 
receive important radiative contributions, their main
contribution is the tree-level one.
From the MSSM minimization equations, $m_A$ is given at tree level by
\begin{equation}
m_A^2 = \left( \frac{t^2_{\beta} + 1}{t^2_{\beta} - 1} \right)
\left(m_{H_d}^2 - m_{H_u}^2 \right) - M_Z^2.
\end{equation}
The fixed point predicts $h_t(M_G) > h_b(M_G)$ and
\begin{equation}
m_{H_d}^2(M_G) - m_{H_u}^2(M_G) = \frac{109}{521} M_{1/2}^2 >0.
\end{equation}
The structure of the MSSM RGEs keeps the difference 
$(m_{H_d}^2 - m_{H_u}^2)$ positive from the GUT to the $M_Z$ scale.
As a consequence, a satisfactory tree-level electroweak symmetry breaking
is possible if the gaugino mass is large enough. 
In Fig.~\ref{fig:su5spectra}
we plot the spectra as a function of the neutralino mass.
Numerically we find that 
in order to
achieve electroweak symmetry breaking
we must have $m_{\chi^0_1} > 50$~GeV,
corresponding to the constraint $M_{1/2} \gtrsim 145$~GeV.
The region excluded by this constraint is shown
by the dark shaded (red) area 
in Fig.~\ref{fig:su5spectra}. 
On the other hand, the experimental measurement of the 
tau lepton mass implies that 
the fixed point is phenomenologically viable only when $\tan\beta$
is large. At such a limit the $\mu$-term is given at tree level by
\begin{equation}
\mu^2 \simeq - m_{H_u}^2 - \frac{1}{2}M^2_Z >0.
\end{equation}
Therefore, $\mu^2$ is automatically positive if 
$M_{1/2}$ (and hence -$m_{H_u}^2$) is large enough,
because the fixed point predicts $m_{H_u}^2<0$.
This bound on $M_{1/2}$ is milder than the 
constraint derived from the condition $m^2_A>0$.
Since the soft parameters are proportional to $M_{1/2}$
, sparticle masses are to a good approximation also
proportional to the unified gaugino mass.
This characteristic can be clearly observed in Fig.~\ref{fig:su5spectra}.
For a neutralino mass $m_{\chi^0_1} \lesssim 100$~GeV,
the fixed point predicts  a
Higgs boson experimentally excluded, $m_h <114$~GeV
(see grey area in Fig.~\ref{fig:su5spectra}).
The Higgs boson masses have been computed including 
one loop sfermion corrections \cite{MSSMeffective},
while the other masses have been computed at tree level.
As can be seen in Fig.~\ref{fig:su5spectra},
a clear mass pattern arises, 
the SUSY-QCD sector being the heaviest:
$$
m_{\widetilde{g}} > m_{\widetilde{t}_2} \simeq 
m_{\widetilde{b}_2} > m_{\widetilde{b}_1} \gtrsim m_{\widetilde{t}_1}.
$$
The lightest neutralino, $m_{\chi^0_1}$, is mostly B-ino.
The heavier chargino and the two heaviest neutralinos are mostly
Higgsinos, and form an approximately degenerate singlet, $\widetilde{\chi}^0_3$,
and triplet, 
$(\widetilde{\chi}^+_2, \widetilde{\chi}^0_4, \widetilde{\chi}^-_2)$,
of ${\rm SU(2)}_L$, all with masses set by the $\mu$-parameter.
The second lightest neutralino and the lightest chargino
are mostly W-inos and form a degenerate triplet of ${\rm SU(2)}_L$,
$(\widetilde{\chi}^+_1, \widetilde{\chi}^0_2, \widetilde{\chi}^-_1)$.
The following mass pattern arises:
$$
m_{\widetilde{\chi}^0_{3,4}} \simeq m_{\widetilde{\chi}^{\pm}_2}
> m_{\widetilde{\tau}_2} \simeq m_{\widetilde{\nu}_{\tau}} >
m_{\widetilde{\tau}_1} > m_{\widetilde{\chi}^0_2} \simeq  
m_{\widetilde{\chi}^{\pm}_1}
> m_{A} \simeq m_H \simeq m_H^{\pm}.  
$$
The fixed point does not only predict qualitative relations between
the masses of the supersymmetric spectra. 
Except for $m_h$, 
the ratios of masses are 
more or less fixed, as shown in Fig.~\ref{fig:ratsu5spectra}:
\begin{eqnarray}
\left( m_{\widetilde{g}}, m_{\widetilde{t}_2, \widetilde{b}_2},
m_{\widetilde{b}_1}, m_{\widetilde{t}_1} \right)
&\simeq& \left( 5.5,~4.7,~4.4,~4.2 \right) \times m_{\widetilde{\chi}^0_1}, 
\label{sqcdratios}
\\
\left( m_{\widetilde{\chi}^{\pm}_2,\widetilde{\chi}^{0}_{3,4}}, 
m_{\widetilde{\tau}_2, \widetilde{\nu}_{\tau}},
m_{\widetilde{\tau}_1}, 
m_{\widetilde{\chi}^{\pm}_1,\widetilde{\chi}^{0}_{2}} \right)
&\simeq& \left( 2.9,~2.5,~2.2,~1.8 \right) \times m_{\widetilde{\chi}^0_1},
\\
m_{A,H,H^{\pm}} &\simeq& 1.34 \times m_{\widetilde{\chi}^0_1}.
\end{eqnarray}
%
Since $m^2_{\tilde{t}_L} 
\simeq m^2_{\tilde{t}_R} \simeq m^2_{\tilde{b}_R}$
and $m^2_{\tilde{\tau}_L} 
\simeq m^2_{\tilde{\tau}_R}$, the diagonal entries
of the sfermion matrices are very similar, and  
the sfermion mixing is quasimaximal. 
%

In summary, the fixed point (a) is compatible with a 
correct electroweak symmetry breaking if $M_{1/2} \gtrsim 150$~GeV, (b) predicts
the neutralino as the LSP for all the allowed
parameter space, and (c) predicts definite relations between the 
masses of the superspectra (see Fig.~\ref{fig:ratsu5spectra}). 
These mass relations 
could give us a clear
confirmation of the SU(5) fixed point scenario in future experiments, once several
supersymmetric particles have been observed. 
%
\section{top and bottom mass predictions including one-loop 
supersymmetric thresholds\label{sec6}}
%
\begin{figure}
\begin{center}
\includegraphics[angle=0, width=0.8\textwidth]{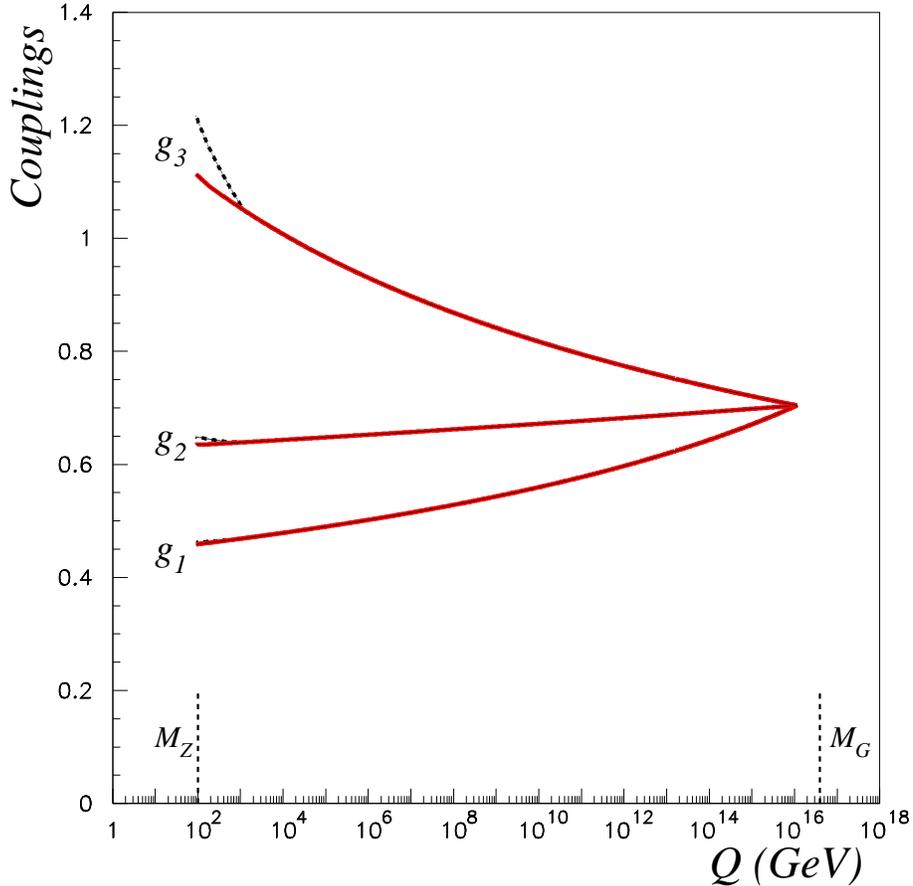}
\caption{\rm Gauge coupling 
running at two-loop (solid red line) in the MSSM from the GUT to the $M_Z$ 
scale, plus one-loop logarithmic SUSY 
thresholds (black dashed line). The 
plot corresponds to the point 1 in Table~\ref{su5pointsmun}.}
\label{fig:su5thryuk}      
\end{center}
\end{figure}      
%
Once we know the supersymmetric spectra predicted by the SU(5)
fixed point, we can refine the predictions for gauge couplings
and for top and bottom quark masses, including the low-energy 
supersymmetric threshold corrections.
Incorporating these is the main purpose of this section.
We begin with a description of our numerical procedure for 
implementing these thresholds. 
%
\subsection{Low-energy supersymmetric thresholds}
%
Assuming a common decoupling scale for all the 
supersymmetric particles, the logarithmic SUSY threshold 
corrections to the strong coupling constant $\alpha_s(M_Z)$
would be given by a simple formula
\cite{loggaug}:
\begin{equation}
\alpha_s^{\rm SM}(m_Z)= \frac{\alpha_s^{\rm MSSM}(m_Z)}{1- 
\frac{2\alpha_s^{\rm MSSM}(m_Z)}{\pi}\ln\left(\frac{M_{\rm SUSY}}{m_Z}\right)} > 
\alpha_s^{\rm MSSM}(m_Z).
\label{gaugethres}
\end{equation}
%
We observe that in general the supersymmetric thresholds 
increase the value of $\alpha^{\rm SM}_s(M_Z)$ 
with respect to the MSSM prediction.
In our study, the logarithmic thresholds to the three
gauge couplings have been implemented
numerically, decoupling one by one the supersymmetric particles
from the one-loop beta functions in the integration of the 
renormalization group equations from the GUT to the $M_Z$ scale.
The effect of the supersymmetric
thresholds on the gauge couplings
is shown in Fig.~\ref{fig:su5thryuk} (dashed black line)
for one representative point of our scans.
The solid-red line corresponds to the two-loop MSSM gauge coupling 
running with no thresholds.
The effect of the thresholds is crucial
for a precise prediction of $\alpha_s$, 
$\widehat{\alpha}_{em}$ and 
$\widehat{s}^2_W$
\cite{gaugeunif,loggaug}.
The thresholds to $\alpha_1$ and $\alpha_2$
are smaller but no
less important, due to the small uncertainty in their experimental
measurement.

The physical top quark mass is calculated from the expression
\begin{equation}
m_t = h_t(m_t) v(m_t) s_{\beta} \left( 1 + \Delta_{\rm QCD}^{\rm t} 
+ \Delta_{\rm SUSY}^{\rm t}
\right),
\end{equation}
where $\Delta_{\rm QCD}^{\rm t}$ stands for the gluon correction that 
was given in Eq.~(\ref{gluoncor}), 
and $\Delta_{\rm SUSY}^t$ is the
supersymmetric contribution evaluated at $m_t$ scale. 
$\Delta_{\rm SUSY}^t$ includes 
contributions from all the third generation 
MSSM spectra. We have implemented the complete
one-loop contributions to $\Delta_{\rm SUSY}^t$ as 
given in Ref.~\cite{Pierce:1996zz}. {\it A posteriori}, we observe that
the dominant contribution to $\Delta^t_{\rm SUSY}$ is the logarithmic
component of the gluino-stop diagrams, 
$\Delta^t_{\widetilde{g}-{\rm Log}}$,
which is in general positive for $m_{\widetilde{g}} > m_t$. 
This contribution is
\begin{equation}
\Delta^t_{\widetilde{g}-{\rm Log}} = \frac{1}{3}\frac{\alpha_s(m_t)}{\pi}
\left[ \ln\left(\frac{m_{\widetilde{g}}^2}{m^2_t}\right) - \frac{1}{2} 
+ \sum_{i=1,2} g(x_{\widetilde{t}_i}) \right],
\label{toplogcor}
\end{equation}
%
where $x_{\widetilde{t}_{1,2}} 
= m^2_{\widetilde{t}_1, \widetilde{t}_2}/m^2_{\widetilde{g}}$, and the function 
$g(x)$ is given by
\begin{equation}
g(x) = 
\frac{1}{2} \left[\ln x - \frac{1}{1-x} - \frac{\ln x}{(1-x)^2}\right].
\end{equation}
This function satisfies $\left| g(x) \right| < 1/2$ if $0<x<1$ and
$g(1) = 1/4$.
At the fixed point, using the
predicted ratios for the SUSY spectra given in Eq.~(\ref{sqcdratios}),
a simple expression for 
this contribution can be written, 
which works as a very good approximation: 
\begin{equation}
\left( \Delta^t_{\widetilde{g}-{\rm Log}} \right)_{{\rm SU(5) f.p.}} 
\simeq \frac{1}{3}\frac{\alpha_s(m_t)}{\pi}
\left[ \ln\left(\frac{m_{\widetilde{g}}^2}{m^2_t}\right) - \frac{1}{4}  \right].
\label{toplogapr}
\end{equation}
%
The standard model running bottom quark
mass just above $m_Z$ in the $\overline{\rm DR}$ scheme is computed using
the formula
\begin{equation}
m_{b}^{\overline{\rm DR}}(m_Z^>) = h_{b}(m_Z) v(m_Z) \cos\beta
\left( 1 + \Delta_{\rm SUSY}^b \right),
\end{equation}
where $m_{b}^{\overline{\rm DR}}(m_Z^>)$ is related to $m_{b}^{\overline{\rm MS}}(m_Z^<)$,
the running mass just below $m_Z$,  
as explained in Sect.~\ref{sec2}.
 $\Delta_{\rm SUSY}^b$ is the
supersymmetric threshold evaluated at the $m_Z$ scale. 
$\Delta_{\rm SUSY}^b$ includes contributions 
from all the third generation 
MSSM spectra. We have implemented the complete
one-loop expressions given in Ref.~\cite{Pierce:1996zz}.
We observe that there are three dominant contributions to 
$\Delta_{\rm SUSY}^b$:
the logarithmic, $\Delta^b_{\widetilde{g}-{\rm Log}}$,
and finite, $\Delta^b_{\widetilde{g}-{\rm Fin}}$, pieces
of the gluino-sbottom diagram
and the chargino-stop contribution, 
$\Delta^b_{\widetilde{\chi}-\widetilde{t}}$. 
These are given below.
The logarithmic part of
the gluino-sbottom contribution is analogous to the
gluino-stop contribution to the top mass:
\begin{equation}
\Delta^b_{\widetilde{g}-{\rm Log}} = \frac{1}{3}\frac{\alpha_s(m_Z)}{\pi}
\left[ \ln\left(\frac{m_{\widetilde{g}}^2}{m^2_Z}\right) - \frac{1}{2} 
+ \sum_{i=1,2} g(x_{\widetilde{b}_i})
\right].
\label{botlogcor}
\end{equation}
%
At the fixed point, there is a
simpler formula analogous to Eq.~(\ref{toplogapr})
\begin{equation}
\left( \Delta^b_{\widetilde{g}-{\rm Log}} \right)_{{\rm SU(5) f.p.}} 
\simeq \frac{1}{3}\frac{\alpha_s(m_Z)}{\pi}
\left[ \ln\left(\frac{m_{\widetilde{g}}^2}{m^2_Z}\right) - 
\frac{1}{4}  \right],
\label{botlogapr}
\end{equation}
%
which also works as a very good approximation.
The finite part of the gluino-sbottom contribution is given at large 
$\tan\beta$ by
\begin{equation}
\Delta^b_{\widetilde{g}-{\rm Fin}} = \frac{2}{3}\frac{\alpha_s(m_Z)}{\pi}
\frac{\mu t_{\beta}}{m_{\widetilde{g}}}
f(x_{\widetilde{b}_1}, x_{\widetilde{b}_2}),
\label{botfincor}
\end{equation}
%
with 
$x_{\widetilde{b}_{1,2}}= m^2_{\widetilde{b}_{1,2}}/m_{\widetilde{g}}^2$ and
$m_{\widetilde{b}_{2}} > m_{\widetilde{b}_{1}}$.
The function $f(x,y)$ is given by
\begin{equation}
f(x,y) = 
\frac{\left[(1- y) x \ln x -
(1- x ) y \ln y \right]}
{(y - x)(1-x)(1-y)},
\end{equation}
%
where $f(x_{\widetilde{b}_1},
x_{\widetilde{b}_2})>0$.
Therefore, the sign of $\Delta^b_{\widetilde{g}-{\rm Fin}}$ 
is the sign of the
$\mu$-term. Moreover, the ratios $x_{\widetilde{b}_{1,2}}$ are well 
determined by the
SU(5) fixed point predictions, $x_{\widetilde{b}_{1}} \simeq 0.61$ and
$x_{\widetilde{b}_{1}} \simeq 0.74$. 
We thus obtain $f(x_{\widetilde{b}_1},
x_{\widetilde{b}_2})\simeq 16/25$, so that 
to a very good approximation,
\begin{equation}
\left( \Delta^b_{\widetilde{g}-{\rm Fin}} \right)_{{\rm SU(5) f.p.}}
\simeq \frac{32}{75}\frac{\alpha_s(m_Z)}{\pi}
\frac{\mu t_{\beta}}{m_{\widetilde{g}}}.
\label{finbotcor}
\end{equation}
%
At large $\tan\beta$,
the dominant contribution to the chargino-stop diagram is given by
\begin{equation}
\Delta^b_{\widetilde{\chi}-\widetilde{t}} \simeq \frac{h^2_t}{16\pi^2}
\frac{A_t t_{\beta}}{\mu}
f(x_{\widetilde{t}_1}, x_{\widetilde{t}_1}),
\label{charbotcor}
\end{equation}
%
where $x_{\widetilde{t}_{1,2}}= m^2_{\widetilde{t}_{1,2}}/\mu^2$. The
fixed point predicts
$m_{\widetilde{t}_{2}} > m_{\widetilde{t}_{1}} > \left| \mu \right|$
and, in general, $A_t<0$. Thus, the sign of this contribution 
is inversely correlated with the sign of $\mu$.
The standard model running tau
mass just above $m_Z$ is computed using
the formula
\begin{equation}
m_{\tau}(m_Z^>) = h_{\tau}(m_Z) v(m_Z) \cos\beta
\left( 1 + \Delta_{\rm SUSY}^{\tau} \right),
\label{tauthres}
\end{equation}
where $m_{\tau}(m_Z^>)$ is related to the
running mass just below $m_Z$ as explained in Sect.~\ref{sec2}.
 $\Delta_{\rm SUSY}^{\tau}$ implements the
complete one-loop supersymmetric threshold evaluated at the $m_Z$ scale
\cite{Pierce:1996zz}.
We observe that there is a dominant contribution to 
$\Delta_{\rm SUSY}^{\tau}$,
the chargino loop given by
\begin{equation}
\Delta^{\tau}_{\widetilde{\chi}^{\pm}-\widetilde{\nu}_{\tau}} = 
- \frac{\alpha_2(m_Z)}{4\pi}\frac{M_2 t_{\beta}}{\mu}
f(x_{M_2}, x_{\widetilde{\nu}_{\tau}}),
\label{taucharcor}
\end{equation}
%
where $x_{M_2}= M_2^2/\mu^2$ and $x_{\widetilde{\nu}_{\tau}}= 
m^2_{\widetilde{\nu}_{\tau}}/\mu^2$. As a result,
$f(x_{M_2}, x_{\widetilde{\nu}_{\tau}})>0$,  and
$\Delta^{\tau}_{\widetilde{\chi}^{\pm}-\widetilde{\nu}_{\tau}}$ 
is inversely correlated with the sign of $\mu$.
The fixed point predicts
$\left| \mu \right| > m_{\widetilde{\nu}_{\tau}} > M_2$, 
and since $\alpha_2(m_Z)$ is approximately $0.03$,
we expect this contribution to be much smaller than the chargino-stop
contribution to the bottom mass. We obtain typically 
$\Delta^{\tau}_{\widetilde{\chi}^{\pm}-\widetilde{\nu}_{\tau}}\simeq +
4\%$ to $+ 6\%$ for $\mu<0$ and $-4\%$ to $- 6\%$ for $\mu>0$, 
which turns out to be important for a precise determination of $\tan\beta$.
%
\subsection{Mass spectra: numerical results}
%
The complete SU(5) fixed point
scenario contains four parameters: $M_G$, $g$, $M_{1/2}$, and $\tan\beta$.
The numerical procedure that we will use in this case is 
a generalization of the procedure we used in Sect.~\ref{sec2} for the 
case with no supersymmetric thresholds. 
We first scan the model parameter space $(M_G, g)$ for
different values of $M_{1/2}$ and integrate numerically 
the complete set of two-loop MSSM RGEs from the GUT to the $m_Z$ scale. 
We next compute the low-energy supersymmetric 
spectra and supersymmetric thresholds to gauge couplings, and
consider only those solutions that
fix the experimental values of $\alpha_s(m_Z)_{\overline{\rm MS}}$ 
and $\alpha_e^{-1}(m_Z)_{\overline{\rm MS}}$. 
Then we use the measured value of the tau mass 
and Eq.~(\ref{tauthres}) to fix $\tan\beta$.
Once we know $\tan \beta$, we can compute the thresholds to the fermion masses. 
Finally, we obtain three predictions for $m_t$, 
$m_b(m_Z)_{\overline{\rm MS}}$, and 
$\widehat{s}^2_W(m_Z)_{\overline{\rm MS}}$. 

Since threshold corrections to fermion masses depend on the sign
of $\mu$, we must distinguish two cases: $\mu<0$, for which 
results are shown in Table~\ref{su5tablemun}, and 
$\mu>0$, for which results are shown
in Table~\ref{su5tablemup}. 
We begin with 
four choices of $M_{1/2}$: $250, 500, 1000$, and $1500$~GeV. 
We show in Tables~\ref{su5tablemun} and \ref{su5tablemup} our 
theoretical predictions for 
$\widehat{s}^2_W(M_Z)_{\overline{\rm MS}}$, 
$m_b(m_Z)_{\overline{\rm MS}}$, $m_t$,
the Higgs boson mass $h$, and the LSP (lightest supersymmetric
particle), along with the values of the input parameters
$M_G$, $g$ and $\tan\beta$. 
The following points are worth noting
\begin{itemize}
\item
The top mass is predicted
to be well within the experimental range, $m_t \simeq 174.3 \pm 5.1$~GeV.
Moreover, its predicted value
turns out to be only weakly sensitive to $M_{1/2}$
and the sign of $\mu$,
ranging between $174.4$~GeV and $176.9$~GeV,
as shown in Fig.~\ref{fig:topgut}.
\item
It is well known that 
predictions for the weak mixing angle in supersymmetric
unified models, with SUSY spectra below $2$~TeV, 
are not completely successful but still very good.
It is important to keep in mind that 
according to PDG $\widehat{s}^2_W(M_Z)_{\overline{\rm MS}}=
0.23117(16)$, thus $\sigma$, the experimental uncertainty, is only 
$0.07\%$. Our model predicts that  
for $M_{1/2}=250$~GeV, the central value for 
$\widehat{s}^2_W(M_Z)_{\overline{\rm MS}}$ is $0.2338$,
which is more than $+16~\sigma$ away from the experimental
value; and for $M_{1/2}=1.5$~TeV, our central value is
$0.2322$ which is at $+6~\sigma$. In spite of this,
it must be considered a good prediction since, if we take into 
account that the experimental uncertainty,  
we predict the weak mixing angle correctly below $1\%$ precision.
\item 
The prediction for the bottom mass is very sensitive to
the sign of $\mu$ and to the unified gaugino mass.
For $\mu>0$, our predicted value is clearly too large.
For $\mu<0$, it is uncomfortably small but gets close
to its experimental value,
$m_b^{\overline{\rm MS}}(m_Z^>)=2.86 \pm 0.2$~GeV, 
for large values of the gaugino mass, as shown in Fig.~\ref{fig:botgut}.
We will focus on $\mu<0$ below.
\item
The higher the gaugino unified mass, the lower
the gauge unified coupling, and the lower
the unification scale. This is because the
thresholds on the value of the gauge
couplings at $m_Z$, are always
positive, as can be seen in Eq.~(\ref{gaugethres}) and 
Fig.~\ref{fig:su5thryuk},  
and they increase with the gaugino mass. Numerically, we find that 
for $\alpha_s$ the threshold corrections are $\sim 19 \%$.
\end{itemize}
%
\begin{table} \centering    
\begin{tabular}{|c|c|c|c|c|}
\hline \hline
  parameters   & A &  B &  C &  D  \\ \hline
$M_{1/2}$~(GeV) &  $250$ &  $500$ &  $750$  & $1500$  \\ \hline 
\multicolumn{5}{|c|}{experimental constraints} \\ \hline
$\alpha_e(m_Z)^{-1}_{\overline{\rm MS}}$ & \multicolumn{4}{c|}{$127.934\pm 0.027$}  \\ 
$\alpha_s(m_Z)_{\overline{\rm MS}}$ &  \multicolumn{4}{c|}{$0.1172 \pm 0.0020$}  \\ 
$m_{\tau}$~(GeV) & \multicolumn{4}{c|}{$1.77703$}  \\ \hline
\multicolumn{5}{|c|}{model parameters} \\ \hline
$M_{G}\times 10^{-16}$~(GeV) & $1.205 \pm 0.139$ 
& $1.130 \pm 0.126$   
& $1.074 \pm 0.020$  & $0.9995 \pm 0.1122$  \\
$\alpha_G^{-1}$ & $0.7094 \pm 0.001$ 
& $0.7037 \pm 0.0011$    
& $0.7005 \pm 0.0011$ & $0.6951 \pm 0.0011$ \\ 
$\tan \beta$ & $51.74 \pm 0.03$ & $51.626 \pm 0.02$ & 
$51.62 \pm 0.02$ & $50.569 \pm 0.056$ \\  \hline
\multicolumn{5}{|c|}{theoretical predictions} \\ \hline
$s_W^2(m_Z)_{\overline{\rm MS}}$ & $0.2338 \pm 0.0005$ & $0.2332 \pm 0.0005$ & 
$0.2328 \pm 0.0005$  & $0.2322 \pm 0.0005$ \\
$m_t$~(GeV) & $176.01 \pm 0.97$ &  $175.67 \pm 1.06$ & 
$175.39 \pm 1.08$ & $174.401 \pm 1.06$ \\ 
$m_b(m_Z)_{\overline{\rm MS}}$~(GeV) & $2.023 \pm 0.005$  & $2.159 \pm 0.006$ 
& $2.215 \pm 0.008$ &  $2.376 \pm 0.010$ \\ 
$m_h $~(GeV) & $ 113.85 \pm 0.14$  
& $119.09 \pm 0.06$ & 
$121.32 \pm 0.04$  &  $123.362 \pm 0.03$ 
\\ 
$m_{\chi^0_1}$~(GeV) & $101.43 \pm 0.47$  & $206.67 \pm 0.92$ & 
$312.58 \pm 1.35$  &  $632.98 \pm 2.74$ \\
\hline \hline
\end{tabular} 
  \caption{\rm Predictions from a minimal supersymmetric
SU(5) exact fixed point, 
including low-energy supersymmetric threshold corrections,
for $M_{1/2}<1.5$~TeV and $\mu<0$.}
  \label{su5tablemun}    
\end{table} 
%
%
\begin{table} \centering    
\begin{tabular}{|c|c|c|c|c|}
\hline \hline
  parameters   & A &  B &  C &  D  \\ \hline 
$M_{1/2}$~(GeV) &  $250$ &  $500$ &  $1000$  & $1500$  \\ \hline 
\multicolumn{5}{|c|}{experimental constraints} \\ \hline
$\alpha_e(m_Z)^{-1}_{\overline{\rm MS}}$ & \multicolumn{4}{c|}{$127.934\pm 0.027$}  \\ 
$\alpha_s(m_Z)_{\overline{\rm MS}}$ &  \multicolumn{4}{c|}{$0.1172 \pm 0.0020$}  \\ 
$m_{\tau}$~(GeV) & \multicolumn{4}{c|}{$1.77703$}  \\ \hline
\multicolumn{5}{|c|}{model parameters} \\ \hline
$M_{G}\times 10^{-16}$~(GeV) & $1.205 \pm 0.139$ 
& $1.130 \pm 0.126$   
& $1.074 \pm 0.121$  & $0.9995 \pm 0.112$  \\
$\alpha_G^{-1}$ & $0.7094 \pm 0.001$ 
& $0.7038 \pm 0.0011$    
&  $0.700 \pm 0.001$ & $0.6951 \pm 0.001$ \\ 
$\tan \beta$ & $43.727 \pm 0.015$ & $43.460 \pm 0.012$ & 
$44.05 \pm 0.012$ & $43.836 \pm 0.02$ \\  \hline
\multicolumn{5}{|c|}{theoretical predictions} \\ \hline
$s_W^2(m_Z)_{\overline{\rm MS}}$ & $0.2338 \pm 0.0005$ & $0.2332 \pm 0.0005$ & 
$0.2328 \pm 0.0005$ & $0.2322 \pm 0.0005$ \\
$m_t$~(GeV) & $176.03 \pm 0.97$ &  $175.70 \pm 1.06$ & 
$175.14 \pm 1.08$ & $174.42 \pm 1.06$ \\ 
$m_b(m_Z)_{\overline{\rm MS}}$~(GeV) & $4.436 \pm 0.04$  & $4.335 \pm 0.04$ 
& $4.216 \pm 0.040$ &  $4.139 \pm 0.038$ \\ 
$m_h $~(GeV) & $ 114.56 \pm 0.11$  
& $119.40 \pm 0.073$ & 
  $121.58 \pm 0.04$  &  $123.55 \pm 0.03$ 
\\ 
$m_{\chi^0_1}$~(GeV) & $101.01 \pm 0.45$  & $206.30 \pm 0.92$ & 
$312.443 \pm 1.349$  &  $632.256 \pm 2.63$\\ 
\hline \hline
\end{tabular} 
\caption{ \rm 
Predictions from a minimal supersymmetric
SU(5) exact fixed point, 
including low-energy supersymmetric threshold corrections,
for $M_{1/2}<1.5$~TeV and $\mu>0$.}
  \label{su5tablemup}    
\end{table} 
%
%
\subsection{Predictions for multi-TeV gaugino mass and naturalness}
%
In light of the successful top-quark-mass prediction of
the SU(5) fixed point scenario,
one is compelled to consider extreme possibilities, which,
surprising as it may seem, could
allow the fixed point to predict correctly the bottom mass
and the weak mixing angle.
The alert reader may have realized (see Table~\ref{su5tablemun})
that for large gaugino unified mass and $\mu<0$,  
not only does the bottom mass approach its experimental value, but the 
prediction for the weak mixing angle also approaches its
experimentally measured value.

\begin{figure}
\begin{center}
\includegraphics[angle=0, width=0.8\textwidth]{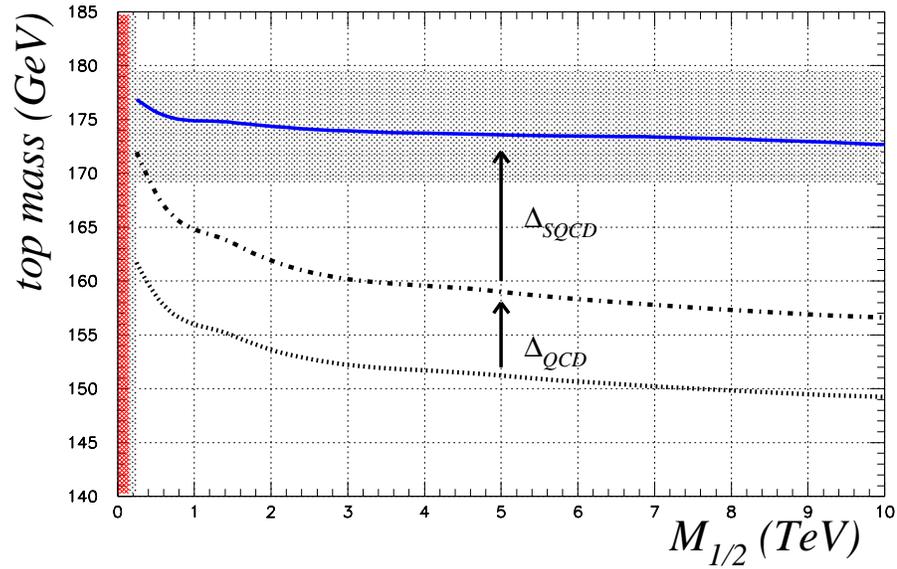}
\caption{\rm 
Top mass prediction
from the minimal supersymmetric SU(5)
exact fixed point with $\mu<0$ as a function of the gaugino mass. The
shaded area represents the experimental world average.}
\label{fig:topgut}
\end{center}
\end{figure}

\begin{figure}
\begin{center}
\includegraphics[angle=0, width=0.8\textwidth]{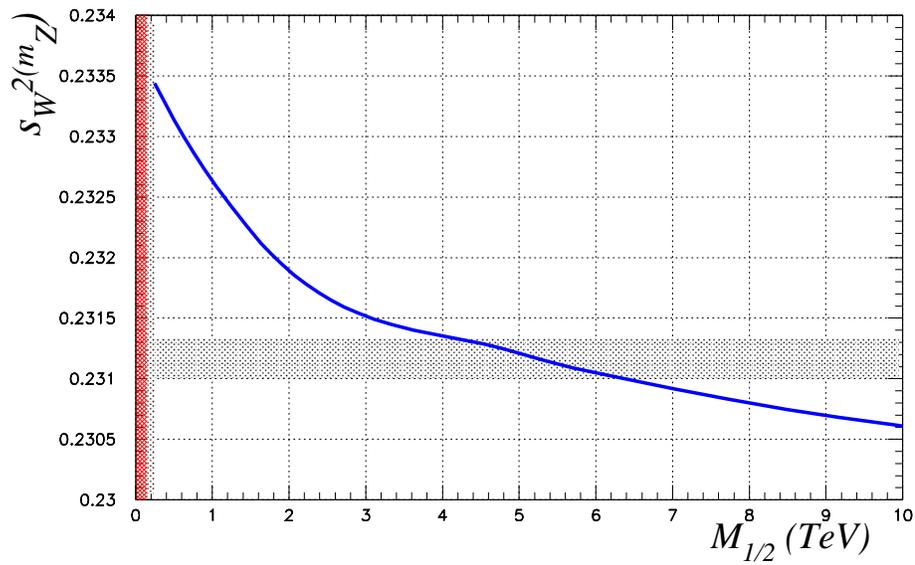}
\caption{\rm
Weak mixing angle prediction
from a minimal supersymmetric SU(5)
exact fixed point, with $\mu<0$, as a function of the gaugino mass. The
shaded area represents the experimentally allowed region.}
\label{fig:sw2gut}
\end{center}
\end{figure}

\begin{figure}
\begin{center}
\includegraphics[angle=0, width=0.8\textwidth]{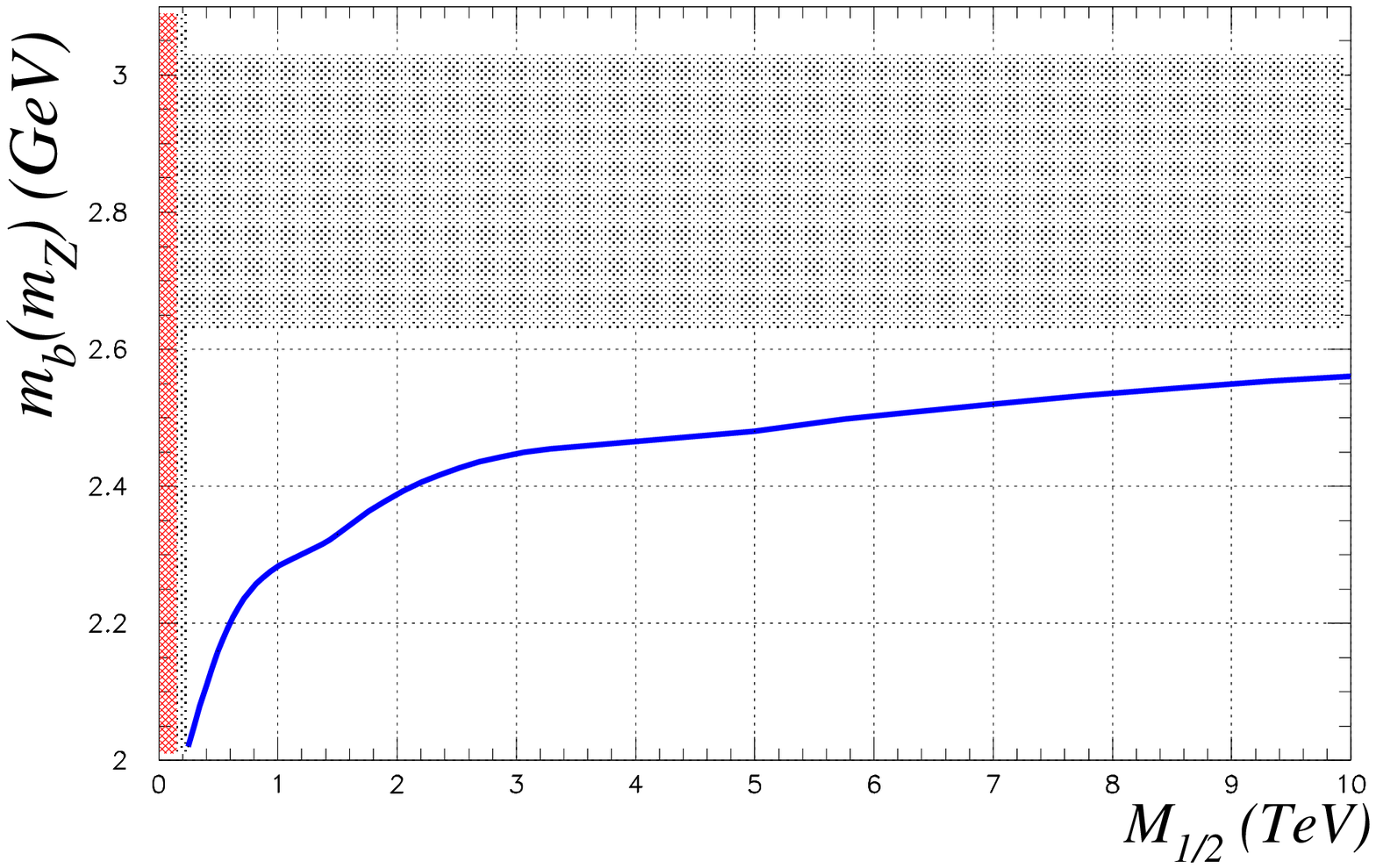}
\caption{\rm   
Bottom mass prediction
from a minimal supersymmetric SU(5)
exact fixed point with $\mu<0$ as a function of the gaugino mass. The
shaded area represents the experimentally allowed region.}
\label{fig:botgut}
\end{center}
\end{figure}

In Table~ \ref{su5pointsmun}, we extend the range of study of
our model to multi-TeV gaugino unified masses.
We show all the parameters and spectra
for five representative points, with $M_{1/2}= 0.5, 1.5, 5, 7$ and $10$~TeV.
We observe that the bottom mass prediction increases continuously
with $M_{1/2}$, going from $2.16$~GeV for $M_{1/2}=500$~GeV
to $2.56$~GeV for $M_{1/2}=10$~TeV, as shown Fig.~\ref{fig:botgut}.
We observe also that the predictions for the weak
mixing angle are good in this range:
the weak mixing angle is $0.23121$ for $M_{1/2}=5$~TeV,
which is inside the
$1~\sigma$ experimental window, $0.23092$ for $M_{1/2}=7$~GeV,
which is inside the $-2~\sigma$ experimental window, and 
$0.23061$ for $M_{1/2}=10$~TeV,
which is inside the $-4~\sigma$ experimental window. 
In Fig.~\ref{fig:sw2gut} we plot the weak mixing angle as
a function of the gaugino mass. 
We also show in Fig.~\ref{fig:topgut} 
the stability of the top mass prediction in this scenario
for very large gaugino masses
as a function of the gaugino mass.
%
%
%
\begin{table} \centering    
\begin{tabular}{|c|c|c|c|c|c|}
\hline \hline
parameters   & point-1 &  point-2 &  point-3 &  point-4  &  point-5 \\ 
\hline 
\multicolumn{6}{|c|}{model parameters} \\ 
\hline
$M_{1/2}$~(TeV) &  $0.5$ &  $1.5$ &  $5$  &  $7$ &  $10$   \\ 
$M_{G}\times 10^{-16}$~({\rm GeV}) &  1.12640  & 0.99135 & 0.86852 & 
0.83742 & 0.81275 \\
$g$ &  0.70386  &  0.69517 & 0.68609 & 0.68359 & 0.68101 \\
$\tan \beta$ &  51.65211   &  51.311 & 51.055 & 50.91695 & 50.74379 \\ 
\hline 
\multicolumn{6}{|c|}{{\rm MSSM Yukawa couplings at $M_G$}} \\
\hline
$h_t(M_G)$ & 0.69406  &  0.68549  & 0.67654  & 0.67408 & 0.67153 \\
$h_b(M_G)$ & 0.53250 &  0.52593 & 0.51906  & 0.51717 & 0.51521 \\
$h_{\tau}(M_G)$ & 0.53250 &  0.52593 & 0.51906 & 0.51717 & 0.51521 \\
\hline
\multicolumn{6}{|c|}{{\rm MSSM dimensionless couplings at $M_Z$}} \\
\hline
$g_1(M_Z)$ & 0.45845   & 0.45659 & 0.45456  & 0.45399 & 0.45337  \\
$g_2(M_Z)$ & 0.63462   & 0.62870 & 0.62241  & 0.62067 & 0.61885 \\
$g_3(M_Z)$ & 1.11336   & 1.07751 & 1.04229  & 1.03298 & 1.02373 \\
$h_t(M_Z)$ & 0.92405  & 0.90163  & 0.87937  & 0.87343 & 0.86745 \\
$h_b(M_Z)$ & 0.80753   & 0.78475  & 0.76220  & 0.75620 & 0.75018 \\
$h_{\tau}(M_Z)$ & 0.48063  & 0.47840 & 0.47575 & 0.47496 & 0.47410 \\
\hline
\multicolumn{6}{|c|}{{\rm experimental constraints}} \\
\hline
$\alpha_e(m_Z)^{-1}_{\overline{\rm MS}}$ & 127.941 & 127.934 & 127.927 & 
127.926 & 127.943 \\ 
$\alpha_s(m_Z)_{\overline{\rm MS}}$ &  0.11727  &  0.11710 & 0.11709 
& 0.11710 & 0.11715 \\ 
$m_{\tau}$~({\rm GeV}) & 1.77703  & 1.77703 & 1.77703 & 1.77703 & 1.77703 \\ \hline
\multicolumn{6}{|c|}{{\rm theoretical predictions}} \\
\hline
$s_W^2(m_Z)_{\overline{\rm MS}}$ & 0.23314 & 0.23222 & 0.23121 & 0.23092 & 0.23061
\\
$m_t$~({\rm GeV}) & 175.720 &  174.732 & 173.580  & 173.379 &  172.674 \\ 
$m_b(m_Z)_{\overline{\rm MS}}$~({\rm GeV}) & 2.157 & 2.330 & 2.481 & 2.519 &
2.561 \\ 
\hline
$\mu$~({\rm GeV}) & -834.757 &  -2273.927 & -6818.149 & -9270.341 &
-12842.62 \\
$B$~({\rm GeV})  & 135.115  &  538.315 & 2210.497 & 3240.261 & 4843.645 \\
\hline 
$h$~({\rm GeV}) & 119.514 & 123.844 & 125.465 & 125.670  & 126.031 \\
$\widetilde{\chi}_1^0$~({\rm GeV}) & 206.620  & 632.977 & 2151.736 
& 3028.433  & 4349.398  \\
$A$~({\rm GeV}) & 334.221 & 979.316 & 3048.760 & 4180.190 &  5838.762 \\
$\widetilde{\chi}_{1}^{\pm}$~({\rm GeV}) & 386.735  &  1181.255 
& 3978.821  & 5584.462 & 7997.872 \\
$\widetilde{\tau}_1$~({\rm GeV}) & 433.172 & 1421.753 & 4793.618 
& 6711.051 & 9584.537 \\
$\widetilde{\nu}_{\tau}$~({\rm GeV}) &  531.169 & 1597.499 & 5307.663 &
7423.032  & 10593.26 \\
$\widetilde{t}_1$~(\hbox{GeV}) &  926.218 &  2762.096  & 8926.256  & 
12376.59 & 17510.36  \\
$\widetilde{b}_1$~(\hbox{GeV}) &  924.958  &  2786.801 &  8983.038
& 12451.25 &  17611.25  \\
$\widetilde{g}$~(\hbox{GeV}) &  1226.465 & 3538.765 & 11348.85 
& 15726.35 & 22241.71 \\
\hline \hline
\end{tabular} 
  \caption{\rm Representative points of the fit for the exact SU(5) fixed point, 
including supersymmetric threshold corrections, for $M_{1/2}<10$~TeV
and $\mu<0$.}
  \label{su5pointsmun}    
\end{table} 
%
%
%
  
Typically, the preservation of naturalness,
caused by quadratic divergences in the radiative corrections to 
the Higgs boson mass,
is assumed to require superpartner masses below a few TeV. 
At present, supersymmetry is the only way we know to 
attack this problem.
Although a rigorous definition of the concept
is lacking, different sensitivity coefficients have been defined 
to measure the fine-tuning degree 
\cite{Ellis:1986yg,Barbieri:1987fn,anderson,Chan:1997bi,strumiafine,deCarlos:yy}.
Some authors have found singularities 
resulting from numerical coincidences
when plotting naturalness limits 
on scalar masses \cite{Barbieri:1987fn,Ross:1992tz}.
This can be interpreted as allowing multi-TeV 
scalar masses \cite{Feng:1999mn}. 

Furthermore, all these criteria for fine-tuning have limitations.
If there were some interrelation between
different parameters caused by some fundamental 
dynamics that leads to the soft terms (such as the presence
of a fixed point), it would show up
in the form of strong correlations between
parameters in the symmetry-breaking solutions;
na\" \i ve fine-tuning
criteria may then be inappropriate for
analyzing the degree of fine-tuning \cite{Carena:1993bs}.
In our SU(5) fixed point scenario, all the soft parameters
are proportional to the gaugino mass, while the Yukawa couplings
are proportional to the unified gauge coupling, and $\tan\beta$
is computed from the tau lepton mass. At tree
level, from the minimization equations, the $Z$ boson mass is 
reinterpreted as a function of the fundamental parameters,
\begin{equation}
m_Z^2 = 2 \left( \alpha_{1/2}M_{1/2}^2 - \mu^2 \right),
\label{finemz}
\end{equation}
where $\alpha_{1/2}$ is a positive constant.
Using a common definition for the fine-tuning sensitivity
coefficients \cite{Feng:1999mn}, one finds that the fine-tuning
of this model can be estimated by the expression
\begin{equation}
c_{\mu} = \left| \frac{\partial \ln m^2_Z}{ \partial \ln \mu} \right|  =
\frac{4 \mu^2}{m_Z^2}.
\label{mucoef}
\end{equation}
For the representative points in Table~\ref{su5pointsmun},
$c_{\mu}$ thus goes from $3\times 10^2$ to $4\times 10^4$.
In light of the successful predictions of this scenario,
we may want to reconsider the use of fine-tuning constraints.

The large value of $c_{\mu}$ suggests 
that this scenario is fine-tuned
so as to reproduce the experimental measurements. 
Furthermore, prospects for 
discovering multi-TeV squarks at the LHC
have been considered \cite{colheavsca}.
From the sample spectra computed for the representative points
in Table~\ref{su5pointsmun} we see that
if the $SU(5)$ fixed point scenario is correct,
it will probably not be possible to discover supersymmetric particles
in the next generation of particle colliders if
the gaugino mass is heavier than 1~TeV.
%
\subsection{Numerical results: threshold corrections}
%
In Tables \ref{pointsradtop}, \ref{pointsradbot}, and
\ref{pointsradtau} 
we include respectively, 
the dominant supersymmetric threshold corrections to top and
bottom quarks and to the tau lepton,
for the five representative points in Table \ref{su5pointsmun}.  
We observe that the logarithmic part of the gluino 
diagram is the dominant contribution to the 
supersymmetric thresholds for the top mass.
This contribution is proportional to 
$\ln \left(m_{\widetilde{g}}/m_t \right)$, and
we find numerically that it varies from $4\%$ (for small $M_{1/2}$)
to $10\%$ (for very large values of $M_{1/2}$). 
Furthermore, we obtain that the prediction for
the top mass is only weakly dependent on the 
gaugino unified mass. This is possible because when we
increase the value of the gaugino unified mass,
the supersymmetric thresholds to gauge couplings also increase,
and as a consequence, we must decrease the unified gauge coupling
to fit the experimental value of $\alpha_s^{\overline{\rm MS}}(m_Z)$. 
When we decrease the unified gauge coupling at the fixed point, 
the top Yukawa coupling decreases (see Table~\ref{su5pointsmun}), and
as a consequence the tree-level component 
of the top quark gets smaller.    
The decrease in the tree-level component is partially counterbalanced
with the increase in the threshold contribution,
and the final prediction for the top pole mass is quite stable.

We now turn to the bottom mass thresholds.
The supersymmetric radiative corrections to $m_b$
for the same representative points of our scans are shown for $\mu<0$ in
Table~\ref{pointsradbot}. 
We see that the dominant correction comes from the
finite piece of the gluino diagram 
and is negative  for $\mu<0$, ranging from 
$-47\%$ (for small $M_{1/2}$) to $-32\%$ (for large $M_{1/2}$).
This negative contribution is partially compensated by
positive corrections caused by the logaritmic part of the
gluino diagram and the chargino-stop loops.
The chargino-stop contribution, Eq.~(\ref{charbotcor}),
is almost independent of $M_{1/2}$ and represents around $14-16$\%,
while the gluino contribution increases with the gaugino mass,
from $+5\%$ (for small $M_{1/2}$) to $+10\%$ (for very large values of $M_{1/2}$).
We observe numerically
that when we increase the gaugino mass, not only do the gaugino
unified coupling and $\tan\beta$ decrease, but the ratio
$\mu/m_{\widetilde{g}}$ also decreases. Therefore, from 
Eq.~(\ref{finbotcor}) we can easily understand 
why the finite part clearly decreases.

With regard to $\tan\beta$, we observe that it decreases
when the gaugino mass increases. For $\mu<0$ we obtain
$\tan \beta=51.65$ for $M_{1/2}=500$~GeV, and
$\tan \beta=50.74$ for very large gaugino mass, $M_{1/2}=10$~TeV.
As explained previously, $\tan\beta$ is determined using the
tau mass. The threshold corrections to the tau mass are inversely
correlated with the sign of $\mu$ and are around $8\%$, as
can be seen in Table~\ref{pointsradtau}. 
These thresholds are crucial because the 
determination of $\tan\beta$ affects the bottom mass significantly.
The inclusion of the tau thresholds represents
a correction to the bottom mass of about $\pm ~(6-7) \%$.
%
\subsection{Theoretical uncertainties} 
%
We will enumerate here the possible theoretical uncertainties
in our predictions.
In Tables~\ref{su5tablemun} and \ref{su5tablemup},
we gave our theoretical predictions including some errors.
Let us clarify that
those errors were due to the uncertainties in the
experimental measurements of $\alpha_s(m_Z)$ and 
$\alpha_e(m_Z)$, which were used as contraints in our scans.
Our numerical results must also contain some theoretical
uncertainties, which we have not yet mentioned. 
These uncertainties would be related to the 
implementation of the radiative corrections
in the low-energy MSSM and to the formulation of the
GUT-scale initial conditions. 
With regard to the low-energy MSSM uncertainties, we note the following 
\begin{itemize}
\item
We have implemented the complete one-loop supersymmetric thresholds
to fermion masses \cite{Pierce:1996zz}. We have seen that
the supersymmetric corrections to 
the bottom quark mass can be very important.
One can wonder if the two-loop supersymmetric corrections 
to the bottom mass
would be crucial for a more precise prediction of the
bottom mass in this scenario.
For instance, 
an additional $3\%$ uncertainty would  
translate into a $\pm 0.1$~GeV theoretical uncertainty.
\item
We have seen above that the leading logarithmic 
supersymmetric threshold corrections to 
the gauge couplings implemented in our analysis \cite{loggaug}
(mainly the strong gauge coupling threshold) 
can be very important, ranging from $+20\%$ to $+30\%$ for large gaugino masses.
It is reasonable to think that the one-loop finite correction
\cite{finitegaug} and perhaps the two-loop SUSY thresholds
may also be important for a more precise prediction 
of $s^2_W(m_Z)$ and $m_b(m_Z)$.
\item
The finite supersymmetric threshold corrections to the bottom mass
are proportional to the $\mu$-term. In our analysis, 
the $\mu$-term is computed from the minimization
equations at $m_Z$, including one loop sfermion corrections
implemented using the effective potential as we
explained in Sect.~\ref{sec5}. 
We have not included the remaining one-loop and
two-loop corrections \cite{Martin:2002iu} to the minimization equations.  
A small percentage of difference in the computation of
the $\mu$-term could significantly affect the
bottom mass prediction.
\end{itemize}
With regard to the GUT scale initial conditions, we note the following
\begin{itemize}
\item
First, we assumed perfect gauge coupling unification.
It is well-known that the GUT-scale threshold
corrections to gauge coupling unification can be
significant \cite{Wright:1994qb,gutthres,Antoniadis:1982vr}. 
These effects would be interesting to include.
\item
We assumed the minimal supersymmetric SU(5) model.
It is probable that we must add extra particle content
to solve the problems that afflict the minimal SU(5) model 
\cite{Masina:2001pp}.
Additional matter content in the unified model would
modify the fixed point predictions. We will estimate
these effects in Sect.~\ref{sec8}.
\item
Finally, we assumed that
the boundary conditions for the MSSM at the unification scale
satisfy the exact SU(5) fixed point predictions.
It would be interesting to analyze how the low-energy predictions
are affected by a small perturbation of the 
fixed point boundary conditions.
\end{itemize}
The successful predictions of our unified fixed point
motivate a more precise study.
The above uncertainties can be seen as a list
of possible improvements if we want to go further 
in the analysis of this scenario.
%
%
%
\begin{table} \centering    
\begin{tabular}{|c|c|c|c|c|c||c|c|}
\hline \hline
  parameters   & $\Delta_{\widetilde{g}-Log}^t$ &
  $\Delta^t_{\widetilde{g}-fin}$  
&  $\Delta_{R}^t$  & $\Delta^t_{\rm SUSY}$ 
& $\Delta_{\rm QCD}^t$ & $h_t(m_t)v(m_t)s_{\beta}$~(GeV) & $m_t$~(GeV) 
\\ \hline 
\hbox{point 1} &  $4.98\%$ &  $0.98\%$ &  $-1.47\%$   & $4.49\%$ & 
$5.98\%$ & 158.67  & 175.72 
\\ \hline
\hbox{point 2} &  $6.68\%$ &  $0.93\%$  &  $-0.77\%$   & $6.84\%$ & 
$5.55\%$ & 154.94 & 174.73 
\\ \hline
\hbox{point 3} &  $8.33\%$ &  $0.87\%$  &  $-0.05\%$   & $9.15\%$ & 
$5.15\%$ & 151.23  & 173.58
\\ \hline
\hbox{point 4} &  $8.87\%$ &  $0.86\%$  &  $0.12\%$   & $9.85\%$ &
$5.04\%$ & 150.23 &  173.38
\\ \hline
\hbox{point 5} &  $9.31\%$ &  $0.84\%$  &  $0.10\%$   & $10.25\%$ &
$4.94\%$ & 149.23 & 172.67 \\
\hline \hline
\end{tabular} 
  \caption{\rm One loop SUSY thresholds to the top 
quark mass evaluated at $m_t$ for the five points 
of Table~\ref{su5pointsmun}. The last column, $m_t$, 
is the theoretical prediction for the pole mass
after adding the SUSY thresholds, $\Delta^t_{\rm SUSY}$,
and the gluon correction, $\Delta_{\rm QCD}^t$, to the tree-level part, 
$h_t(m_t)v(m_t)s_{\beta}$, included in the next-to-last column. 
$\Delta_{\widetilde{g}-Log}^t$ is the dominant  
contribution defined in Eq.~(\ref{toplogcor}), while $\Delta_{R}^t$ includes
the rest of the one-loop SUSY corrections.}
\label{pointsradtop}    
\end{table} 
%
%
\begin{table} \centering    
\begin{tabular}{|c|c|c|c|c|c||c|c|}
\hline \hline
 parameters   & $\Delta_{\widetilde{g}-Log}^b$ 
& $\Delta^b_{\widetilde{g}-Fin}$  
& $\Delta^b_{\widetilde{\chi}^{\pm}}$    
&  $\Delta_{R}^b$  & $\Delta^b_{\rm SUSY}$ 
& $h_b(m_Z)v(m_Z)c_{\beta}$~(GeV) & $m_b^{\overline{\rm MS}}(m_Z^{>})$~(Gev) \\ \hline 
\hbox{point 1} &  $5.31\%$ &  $-46.79\% $ &  
$15.80\%$  & $4.16\%$  & $-21.52\%$  & $2.74$  &  $2.16$ \\ \hline
\hbox{point 2} &  $7.05\%$ &  $-40.91\% $ &  
$15.10\%$  & $5.40\%$  & $-13.36\%$  & 2.68 & $2.33$ \\ \hline
\hbox{point 3} &  $8.73\%$ &  $-35.41\% $ &  
$14.19\%$  & $6.97\%$  & $-5.52\%$ & 2.62 & $2.48$\\ \hline
\hbox{point 4} &  $9.17\%$ &  $-33.98\% $ &  
$13.87\%$  & $7.37\%$  & $-3.57\%$  & 2.61 & $2.52$ \\ \hline
\hbox{point 5} &  $9.62\%$ &  $-32.52\% $ &  
$13.55\%$  & $7.80\%$  & $-1.55\%$ & 2.60 & $2.56$ \\
\hline \hline
\end{tabular} 
  \caption{\rm One-loop SUSY thresholds to the 
bottom quark mass evaluated at $M_Z$ for the five 
points of Table~\ref{su5pointsmun}. The last column, 
$m_b^{\overline{\rm MS}}(m_Z)$, is the theoretical prediction 
for the running mass at $m_Z$ after adding the SUSY thresholds,
$\Delta^b_{\rm SUSY}$, to the tree-level part, $h_b(m_Z)v(m_Z)c_{\beta}$. 
$\Delta_{\widetilde{g}-Log}^b$, 
$\Delta^b_{\widetilde{g}-Fin}$ and $\Delta^b_{\widetilde{\chi}^{\pm}}$ 
are the dominant contributions defined in 
Eqs.~(\ref{botlogcor}), (\ref{botfincor}) and (\ref{charbotcor}) 
respectively, while $\Delta_{R}^b$ includes the remaining one-loop 
SUSY corrections.}
\label{pointsradbot}    
\end{table} 
%
%
\begin{table} \centering    
\begin{tabular}{|c|c|c|c|c||c|c|c|}
\hline \hline
parameters  & $\Delta_{\widetilde{\nu}_{\tau} - 
\widetilde{\chi}^{\pm}}^{\tau}$ 
& $\Delta_{\widetilde{\tau} - 
\widetilde{\chi}^0}^{\tau}$ 
& $\Delta_{R}^{\tau}$ & $\Delta_{\rm SUSY}^{\tau}$ & 
$h_{\tau} (m_Z)v(m_Z)c_{\beta}$ ~(GeV) & 
$m_{\tau}^{\overline{\rm MS}}(m_Z^{>})$~(GeV) & $t_{\beta}$
\\ \hline 
\hbox{point 1} &  $6.26\%$ & $2.02\%$ &  $0.06\%$ &  $8.34\%$ 
& $1.6136$ & 1.7482 & 51.65
\\ \hline 
\hbox{point 2} &  $5.74\%$ & $2.07\%$  &  $0.30\%$ &  $8.11\%$ 
& 1.6169 & 1.7482   &  51.31
\\ \hline
\hbox{point 3} &  $5.35\%$ & $2.23\%$ & $0.59\%$ &  $8.17 \%$   
& 1.6162 & 1.7482 & 51.05
\\ \hline
\hbox{point 4} &  $5.15\%$ & $2.24\%$ & $0.66\%$  &  $8.05\%$   
& 1.6178 & 1.7482 & 50.91
\\ \hline
\hbox{point 5} &  $4.94\%$ & $2.21\%$ & $0.73\%$  &  $7.88\%$   
& 1.6204 & 1.7482 & 50.74 \\
\hline \hline
\end{tabular} 
  \caption{\rm One-loop supersymmetric 
radiative contributions to the tau lepton mass evaluated at $m_Z$ for 
the five representative points of Table~\ref{su5pointsmun}. The last
column, $m_{\tau}(m_Z)$, is the theoretical prediction for the SM 
running mass at $m_Z$ after adding the SUSY thresholds, 
$\Delta_{\rm SUSY}^{\tau}$, to the tree level part, 
$h_{\tau}(m_Z)v(m_Z)c_{\beta}$.
$\Delta_{\widetilde{\nu}_{\tau}-\widetilde{\chi}}^{\tau}$ is the 
dominant contribution defined in Eq.~(\ref{taucharcor}),
while $\Delta_{R}^{\tau}$ includes the remaining one-loop SUSY 
corrections.}
\label{pointsradtau}    
\end{table} 
%
\section{Fixed point predictions in non minimal SU(5) models\label{sec7}}
%
To solve the problems
afflicting the minimal supersymmetric SU(5) model, extended
non minimal models have been examined.
For instance, it is known that one can fit the
first and second generation standard model fermion masses
and mixing angles,
in the context of grand unified models, by adding
extra matter to generate Yukawa textures at the GUT scale.
Many of these extensions contain extra particle
content, suggesting modifications
to the minimal SU(5) fixed point predictions.
Nonetheless,
even though the evolution towards the Yukawa unified fixed point
may occur more rapidly as a result of increased particle content,
the fixed point for the soft parameters would become unstable if the 
added particle content were too large. 
For extensions of the minimal model,
Eq.~(\ref{su5Yukgaurge}) 
applies with the modification
$b=-3 + S$, where $S$ is the sum of Dynkin indices of the
additional fields \cite{dynkin}.
We are lacking a complete
non minimal model for the soft sector and therefore do not 
know exactly the upper bound on $S$ imposed by 
soft stability. A na\" \i ve estimation, however, could be made based on
Eq.~(\ref{su5trilrge}).
\begin{figure}
\begin{center}
\includegraphics[angle=0, width=0.8\textwidth]{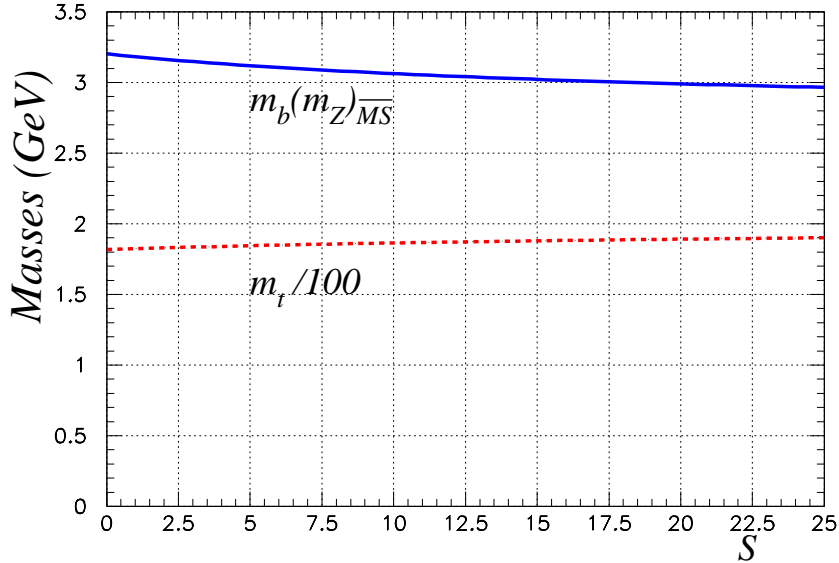}
\caption{\rm 
Top and bottom mass predictions, with no supersymmetric thresholds, 
from a non minimal supersymmetric SU(5)
exact fixed point, with $\mu<0$, as a function of the Dynkin index. 
The value $S=0$ corresponds to minimal supersymmetric SU(5).}
\label{fig:su5nomin}
\end{center}
\end{figure}      
%
The fixed point solution for the Yukawa couplings
in general non minimal models with extra particle content is given by
%
\begin{equation}
\frac{\lambda_t^2}{g^2}  =   \frac{2533}{2605} + \frac{145}{2605}~S,
\qquad
\frac{\lambda_b^2}{g^2}  =   \frac{1491}{2605} + \frac{145}{2605}~S,
\qquad
\frac{\lambda^2_{H}}{g^2}  =   \frac{560}{521} + \frac{30}{521}~S.
\label{su5fpnomin}
\end{equation}
%
In principle $S>0$, so that $\lambda_{H} > 1.036~ g$,
which is a good prediction since 
we overcome more easily the constraints on proton decay.
Although we cannot make precise predictions
for bottom and top masses, we can estimate,
without a complete model of the soft sector of the non minimal
model, how the extra particle content 
affects the mass predictions without including supersymmetric thresholds.
Following the procedure used in Sec.~\ref{sec2},
we study the prediction for the top and bottom mass
as a function of $S$. We show our results in 
Fig.~\ref{fig:su5nomin}. For $S=0$, we recover the
minimal SU(5) predictions.
The bottom mass prediction decreases from $3.2$~GeV for
$S=0$ to $2.95$~GeV for $S=25$,
while the top mass increases from $181$~GeV
to $190$~GeV. 
Surprisingly, the effect of the extra matter is not 
so important as we expected at first sight from 
Eq.~(\ref{su5fpnomin}). 
We see that for $S=0$ the fixed point predicts $h_t = 0.9860~g$,
while for $S=25$ it predicts $h_t = 1.5375~g$,
representing a $56\%$ increase in the GUT scale inital condition.
On the other hand, the top mass prediction increases
only by $5\%$ while the bottom mass decreases by $8\%$. 
This effect is combined with a change in $\tan\beta$
when we increase the particle content. 
The reason is that when the tau Yukawa
coupling grows, $\tan\beta$, which we compute using the
tau mass, also increases. 
The bottom mass is especially affected, compensating for the increase
in the bottom Yukawa coupling.
Even though these conclusions are interesting, they
must be considered preliminary, since non minimal models
imply additional couplings, which affect the RGEs and can modify
these predictions. Furthermore,
we cannot include supersymmetric thresholds without a complete
non minimal model for the soft sector.
\section{Yukawa unification and the large $\tan\beta$ fixed point in the MSSM}
 It seems, from our numerical results, that the SU(5) fixed point
predictions, mainly the top quark mass, are more stable than one might have expected.
The stability of these results could indicate the presence of fixed point 
behavior in the running of the MSSM Yukawa couplings
from the GUT to the $m_Z$ scale.
In the MSSM, there is a fixed point viable at large $\tan\beta$.
The MSSM large $\tan\beta$ fixed point predicts that $h_t^2 = h_b^2 = g_3^2/3$ and $h_{\tau}=0$
in the infrared. The relevance of this fixed point in the context of top--bottom--tau 
unified models was pointed out in Ref.~\cite{Schrempp:1994xn}. 
The large $\tan\beta$ fixed point in the MSSM must not be confused with the low $\tan\beta$ fixed point.
Even though the low $\tan\beta$ fixed point has been much more studied in the
literature \cite{lowQFP}, 
it is not relevant in the context of top-bottom-tau Yukawa unified models
as it predicts that the bottom and tau Yukawa couplings are zero, while 
top-bottom unified models require that the top and bottom Yukawa couplings are of the same 
order at low energies. On the other hand, we think that even though the large $\tan\beta$ fixed point
of the MSSM can be relevant in the SU(5) fixed point scenario, it is not predictive 
enough by itself to explain the measured third generation fermion masses, since
it predicts that the tau Yukawa coupling is zero. 
\section{Summary and comments\label{sec8}}
In this paper we have outlined a precise analysis of
the low-energy implications of a supersymmetric SU(5) fixed point.
It is well known that supersymmetric grand unified theories predict
correctly the weak mixing angle. This unified fixed point
in addition predicts successfully the top quark mass without adjustments to 
any model parameter. This is, to my knowledge, 
the most successful prediction of the top mass in the literature 
based on first principles and implies that $175$~GeV
is a number encoded
in the symmetries of the supersymmetric SU(5) model. Other
interesting predictions of the unified fixed point
studied in this paper are
\begin{itemize}
\item
the bottom mass, which is in general sensitive to the sign of $\mu$ and to the 
gaugino unified mass, approaches its experimental value
for $\mu<0$ and very large values of the gaugino mass;
\item
the weak mixing angle is correctly predicted for low values 
of the gaugino unified mass, as is characteristic
of other supersymmetric unified models, and can be successfully 
predicted for large values of the gaugino unified mass.
\end{itemize}
If the successful prediction of the weak mixing angle 
and the bottom mass (for very large gaugino unified mass) is more 
than a coincidence, it could be a hint that
the mass scale of all supersymmetric particles lies well above 1 TeV.
Furthermore, a very heavy SUSY spectrum would make other
indirect supersymmetric signals unobservable in current experiments.
The results found in this work
may be regarded as evidence for supersymmetry
with grand unification, especially
as grand unification provides a simple and elegant 
explanation of the standard model gauge structure and 
representation content.
%
\section*{Acknowledgments}

I thank Xerxes Tata for many suggestions and constant support.
I thank Tomas Blazek for helpful discussions and
instructive correspondence about the implementation 
of the supersymmetric threshold corrections to fermion masses.
I also thank Hulya Guler for many suggestions. 
This research was
supported in part by the U.S. Department of Energy under grant
DE-FG03-94ER40833.


\end{document}